\documentclass[12pt,letterpaper,draftclsnofoot,onecolumn]{IEEEtran}
\usepackage{amsmath,amsfonts}
\usepackage{amssymb}
\usepackage{algorithmic}
\usepackage{algorithm}
\usepackage{array}
\usepackage[caption=false,font=normalsize,labelfont=sf,textfont=sf]{subfig}
\usepackage{textcomp}
\usepackage{stfloats}
\usepackage{url}
\usepackage{verbatim}
\usepackage{multirow}
\usepackage{booktabs}
\usepackage{physics}
\usepackage{graphicx}
\usepackage{cite}
\newtheorem{theorem}{Theorem}

\newtheorem{lemma}{Lemma}
\newtheorem{remark}{Remark}
\newtheorem{assumption}{Assumption}

\begin{document}

\title{Hopfield Neural Networks for Online\\ Constrained Parameter Estimation with\\Time-Varying Dynamics and Disturbances}
\author{Miguel Pedro Silva\\
Marine Engineering Department\\
Escola Superior Náutica Infante D. Henrique -- ENIDH\\
2770-058 -- Paço de Arcos -- Portugal\\
Email: miguelsilva@enautica.pt}

\maketitle

\begin{abstract}
This paper proposes two projector-based Hopfield neural network (HNN) estimators for online, constrained parameter estimation under time-varying data, additive disturbances, and slowly drifting physical parameters. The first is a constraint-aware HNN that enforces linear equalities and inequalities (via slack neurons) and continuously tracks the constrained least-squares target. The second augments the state with compensation neurons and a concatenated regressor to absorb bias-like disturbance components within the same energy function.

For both estimators we establish global uniform ultimate boundedness with explicit convergence rate and ultimate bound, and we derive practical tuning rules that link the three design gains to closed-loop bandwidth and steady-state accuracy. We also introduce an online identifiability monitor that adapts the constraint weight and time step, and, when needed, projects updates onto identifiable subspaces to prevent drift in poorly excited directions.

A two-degree-of-freedom mass–spring–damper study with Monte Carlo trials compares the proposed HNN estimators against projector-based recursive least squares, disturbance-aware projector-based Kalman filtering, and disturbance-aware projector-based moving-horizon estimation. The HNN estimators achieve competitive or superior accuracy with zero constraint violations, reduced disturbance-induced bias (especially with compensation), and low per-step computational cost suitable for high-rate deployment.
\end{abstract}



\section{Introduction and Background}
\label{sec:intro}

\subsection{Problem setting and motivation}
Online parameter estimation subject to constraints, additive disturbances, and time variation is widespread in engineering: mechanical systems with physical limits (e.g., $m,k,b>0$), robotics with safety envelopes, and process control with operating bounds. In these settings, estimators are sought that (i) enforce linear
equalities/inequalities on parameters, (ii) reject or absorb unmeasured disturbances that corrupt the data, and (iii) track slowly time–varying parameters while remaining computationally light enough for real-time deployment.

To position our approach, we adopt three practitioners' baselines that progressively incorporate these requirements and use them for head-to-head comparisons with the proposed HNN estimators:
\begin{itemize}
\item \emph{PB–RLS (Projection-Based RLS).} A Recursive Least-Squares (RLS) update is followed by a projection onto the feasible set to enforce linear constraints on parameters \cite{MURAKAMI1998,ZHU1999,HASSAN2009,UHLICH2010,WANG2018}. PB–RLS is computationally light and tracks drifts via forgetting, but it does not explicitly model disturbances and typically assumes linear–Gaussian noise.
\item \emph{DA–PB–KF (Disturbance-Augmented, Projection-Based Kalman Filter).} A Kalman filter is augmented with bias/unknown input states to absorb additive disturbances, and each update is projected to satisfy parameter constraints (cf. constrained KF frameworks \cite{LUO2012,JOUKOV2015}). This improves disturbance rejection over PB–RLS at a moderate extra cost, while retaining sensitivity to modelling/non-Gaussian effects.
\item \emph{DA–PB–MHE (Disturbance-Augmented, Projection-Based Moving Horizon Estimation).} A finite-horizon optimisation (MHE) with explicit constraints and disturbance/parameter augmentation, standard in MHE formulations, provides a strong accuracy/robustness baseline under constraints, additive disturbances, and slow parameter drift, albeit with the highest online computational burden (see, e.g., \cite{RAO2001,RAO2003,RAWLINGS2017,ZAVALA2008}).

\end{itemize}

For completeness, we also reference some other widely used general-purpose estimators. Least Mean Squares (LMS) is computationally frugal with effective frequency-domain variants \cite{ARABLOUEI2015,PENG2024}, yet converges slowly and is not inherently constrained \cite{UHLICH2010}. KF/CEKF families achieve Minimum Mean Square Error (MMSE) under Gaussian assumptions and admit constrained projections \cite{LUO2012,JOUKOV2015,HRUSTIC2020,SONG2022}, at higher computational cost and sensitivity to model errors. Particle filters address non-linear/non-Gaussian settings and can encode constraints \cite{PAPI2012,DEMIRBAS2015,HU2023}, but often scale poorly in real time \cite{SATOH2024}. 

In fact, a single estimator that jointly enforces constraints, mitigates disturbances, and tracks parameter drift with clear stability guarantees and low online complexity remains desirable.

Rather than claiming superiority, we propose complementary estimators that map the constrained, disturbance-aware online estimation problem onto a classical Hopfield neural network (HNN) with a projector-based (valid-subspace) approach. The design aims to be competitive in practice by offering an
attractive compromise among: (i) performance under constraints, additive disturbances (via constraint projectors and compensation neurons) and time-varying parameters, (ii) ease of tuning through a small set of interpretable gains $(\alpha,\beta,\eta)$ that directly control bandwidth and ultimate error, and (iii) low online time complexity dominated by matrix–vector products and element-wise activations, which efficiently map to parallel hardware. We evaluate this
trade-off in Section~\ref{sec:simulations} against PB–RLS (constraints only), DA–PB–KF (constraints+disturbances),
and DA–PB–MHE (constraints+disturbances+time-varying parameters).

\subsection{Hopfield neural networks}
Since the seminal work of Hopfield and Hopfield--Tank \cite{HOPFIELD1982,HOPFIELD1985}, analogue recurrent networks have been used both as associative memories and as neurodynamic optimisers, with numerous applications in the 1990s, e.g.,  \cite{TAGLIARINI1991,ALI1993,GEE1994,SMITH1999,ZHANG2014}. Later, high-throughput implementations on GPUs/FPGAs \cite{BASTOSFILHO2011,MEI2014,YAO2014} were proposed. More recently, theoretical developments reinterpret modern Hopfield layers (dense associative memories) as attention-like mechanisms with higher-order energies \cite{KROTOV2016,RAMSAUER2021}. In parallel, differentiable optimisation layers embed convex programmes in deep models \cite{AMOS2017,AGRAWAL2019}, and control-orientated works learn Lyapunov functions and stabilising policies \cite{CHANG2019,DAI2021}.

In contrast, we purposely adopt a classical continuous-time Hopfield Neural Network (HNN) as a lightweight
online estimator and endow it with an energy function construction tailored to constrained and disturbed
streaming data.

The classical HNN with normalised leak/capacitance reads
\begin{subequations}\label{eq:hnn-classic-intro}
\begin{align}
\dot u(t) &= T\,v(t) + b, \\
v(t) &= \alpha\,\tanh\!\Big(\tfrac{\beta}{2}\,u(t)\Big),\qquad v_i\in(-\alpha,\alpha),
\end{align}
\end{subequations}
with neuron gain $\beta>0$, output scaling $\alpha>0$, and (symmetric) weight matrix $T$. For constant
$(T,b)$, trajectories decrease the quadratic energy function
\begin{equation}\label{eq:hnn-energy-intro}
E(v) = -\tfrac12\,v^\top T v - v^\top b + \tfrac12\|b\|^2,
\end{equation}
which acts as a Lyapunov function \cite{HOPFIELD1985,ABE1989}.

A standard way to use an HNN for online Least-Squares (LS) with $w=W\theta \;$ is to identify $v=\theta$ and set
$T=-W^\top W$, $b=W^\top w$. Although simple, this mapping is ill-suited for constrained, time-varying operation:
(i) $T(t),b(t)$ vary with the data, so \eqref{eq:hnn-energy-intro} may no longer be a Lyapunov function;
(ii) $-W^\top W$ can be ill-conditioned, amplifying noise along weakly excited directions; (iii) equality/inequality constraints are not enforced natively. Prior HNN-based parameter estimators \cite{ATENCIA2004,ATENCIA2005,HU2005,ALONSO2007,ALONSO2009,ATENCIA2013,ATENCIA2015,BURAKOV2018} mainly focus on unconstrained problems and do not provide stability guarantees under time variation and disturbances.

In this work, we replace the LS map with a valid-subspace (projector) construction that (i) enforces linear equalities and inequalities (via slack neurons) within the HNN energy function, (ii) absorbs unmeasured additive terms using compensation neurons, and (iii) yields explicit Global Uniform Ultimate Boundedness (GUUB) stability limits when $(T(t),b(t))$ are updated online from the data.

The estimator neurons will represent parameters (and any auxiliary slack/compensator variables). The HNN receives time-varying weights $T(t)$ and bias $b(t)$ synthesised from orthogonal projectors onto the data and constraint subspaces, respectively:
\[
P_W(t)=W(t)^\top\!\big(W(t)W(t)^\top\big)^{-1}W(t),\qquad
P_A=A^\top(AA^\top)^{-1}A,
\]
so that the data part uses $-P_W$ (better conditioned than $-W^\top W$), while constraints contribute $-\eta P_A$ with weight $\eta>0$. Inequalities are lifted to equalities through slack neurons, and unmeasured additive terms $d(t)$ are handled by compensation neurons via $W_{\mathrm{aug}}=[\,W\ H\,]$. The HNN thus minimises a time-varying quadratic energy function composed of (i) data-consistency and (ii) constraint-consistency terms, both expressed through projectors.

\subsection{Contributions}
This work introduces two projector–based HNN estimators for online, constrained
parameter estimation under time-varying data, additive disturbances, and slowly drifting physical parameters:

\begin{enumerate}
\item \emph{Constraint–aware HNN (CA-HNN).} Enforces linear equalities and inequalities (via one slack neuron per inequality) and continuously tracks the constrained least–squares target using a projector mapping in parameter space.
\item \emph{Constraint-aware compensation–augmented HNN (CA$^2$-HNN).} Extends the CA-HNN estimator by adding compensation neurons and a concatenated regressor to absorb bias-like disturbance components within the same energy function.
\end{enumerate}

\noindent The main technical and practical contributions are:

\begin{itemize}
\item \emph{Projector/valid-subspace formulation.} We derive closed-form, time-varying HNN weights from streaming data and fixed constraints via orthogonal projectors in parameter space. This improves conditioning over $W^\top W$, ensures feasibility by design (equalities/inequalities), and makes the dynamics scale-invariant.

\item \emph{Two estimators, one analysis framework.} For the baseline estimator we use the constraint-augmented projector. For the compensation-augmented estimator we use the corresponding augmented projector. Both share the same small set of gains $(\alpha,\beta,\eta)$ and admit the same style of analysis and tuning.

\item \emph{Stability with explicit rate and radius.} We prove constraint-aware contraction and GUUB for both estimators:
Theorem~1 covers the baseline (exogenous disturbances). Theorem~2 covers the compensation-augmented case. In both, the convergence rate depends on an explicit curvature constant, and the ultimate bound separates mapping variation, disturbance power, and parameter drift.

\item \emph{Augmented identifiability for compensation.} We formalise when compensation can separate parameter and disturbance effects (rank/curvature condition on the augmented regressor with constraints acting only on the parameter block).

\item \emph{Practical tuning rules.} We provide simple rules that link the three design gains to closed-loop bandwidth and steady accuracy, and an RK4 step-size guideline tied to the spectral radius of the Jacobian. The same rules apply to the compensated case by replacing the projectors/curvatures with their augmented counterparts.

\item \emph{Online identifiability monitor and mitigation.} We introduce a lightweight scale-invariant score (smallest singular value of a whitened stack) with warning/freeze thresholds. When excitation degrades, the method increases constraint curvature and, if necessary, projects/damps updates along blind directions to prevent drift.

\item \emph{Empirical validation and complexity.} A two-degree-of-freedom mass–spring–damper (2-DOF MSD) study with Monte Carlo trials compares both HNN estimators to PB–RLS, DA–PB–KF, and DA–PB–MHE. The HNN estimators achieve competitive accuracy with zero constraint violations and reduced disturbance-induced bias (especially with compensation), at low online per-step computational cost suitable for parallel implementation.
\end{itemize}

\subsection{Paper organization}
Section~\ref{sec:mapping} introduces the two projector–based HNN estimators and their implementation: 
(i) the constraint–aware baseline (equality and inequality constraints enforced via slack neurons) and 
(ii) the compensation–augmented variant (additional disturbance channel and compensation neurons). 

Section~\ref{sec:stability} states the standing assumptions and rank/curvature lemmas and proves the main results:
Theorem~1 (GUUB for the baseline, exogenous disturbances) and Theorem~2 (GUUB for the compensation–augmented case). 
We make explicit how the gains $(\alpha,\beta,\eta)$ and the constraint–augmented curvature determine convergence rate and ultimate bounds. We also present practical tuning rules (selecting $\alpha$, setting bandwidth via $\beta$, securing curvature with $\eta$, and choosing the RK4 step) and describe an online identifiability monitor with warning/freeze logic and projection/damping mitigation. The same rules apply to the compensated case by substituting the augmented projectors/curvature.

Section~\ref{sec:simulations} reports a 2-DOF MSD study with Monte Carlo runs, 
including quantitative comparisons of the HNN estimators against PB–RLS, DA–PB–KF, and DA–PB–MHE, and an online complexity analysis.

Section~\ref{sec:conclusions} concludes and outlines future work.

\section{Mapping Online Constrained Estimation with Disturbances onto a Hopfield Neural Network}
\label{sec:mapping}

The core idea in this section is to encode linear relations as valid subspaces using
orthogonal projectors, and to derive explicit HNN weights and bias from those projectors.

For any full–row–rank matrix $M$ and right–hand side $m$, the projector onto $\mathrm{range}(M^\top)$ is $P_M=M^\top(MM^\top)^{-1}M$. The affine set of solutions of $Mx=m$
can be written as
\[
x \;=\; T^{\mathrm{val}} x + s, \quad 
\text{where:} \quad T^{\mathrm{val}}=I-P_M,\quad
s=P_M\,x^*\ \ (\text{for any }x^*\ \text{with }Mx^*=m),
\]
so, the quadratic function
\[
E(x)\;=\;\tfrac12\big\|\,x-(T^{\mathrm{val}}x+s)\,\big\|^2
\]
penalises deviations from the valid subspace $\{x:\,Mx=m\}$ \cite{GEE1994}.  In what follows,
we instantiate this construction with $M=W$ (data subspace) and with $M=A$ (constraint subspace,
with inequalities lifted by slack neurons), yielding projector–based Hopfield weights and bias that are
updated online from $(W(t),w(t))$ while enforcing (time-invariant) linear constraints natively.

\subsection{Mapping the unconstrained parameter estimation problem}
Let $w(t)=W(t)\,\theta^*(t)$ and identify the parameter block with the neuron outputs
$v_\theta = \theta$ (whose dimension is $p$). Using $M=W_{q \times p}$ in the construction above gives the data energy function associated to the estimation error
\begin{equation}
E^{\mathrm{ee}}(v_\theta,t)\;=\;-\tfrac12 v_\theta^\top\,T^{\mathrm{ee}}(t)\,v_\theta
\;-\;v_\theta^\top b^{\mathrm{ee}}(t)\;+\;\tfrac12\|b^{\mathrm{ee}}(t)\|^2,
\qquad
T^{\mathrm{ee}}=-P_W,\ \ b^{\mathrm{ee}}=P_W\,v_\theta^*
\label{eq:mapping-ee}
\end{equation}
with $P_W=W^\top(WW^\top)^{-1}W=W^+W$ and $b^{\mathrm{ee}}=W^+w$.
This projector form attenuates ill-conditioned directions compared to least-squares mappings
based on $(W^\top W)^{-1}$.

\subsection{Mapping the constrained (equalities and inequalities) parameter estimation problem}
Equality constraints $A^{\mathrm{eq}}v_\theta=a^{\mathrm{eq}}$ contribute
\[
T^{\mathrm{eq}}=-P_{A^{\mathrm{eq}}},\quad b^{\mathrm{eq}}=P_{A^{\mathrm{eq}}}\,v_\theta^*,\quad
P_{A^{\mathrm{eq}}}=A^{\mathrm{eq}\,\top}(A^{\mathrm{eq}}A^{\mathrm{eq}\,\top})^{-1}A^{\mathrm{eq}}
\]
Inequalities $A^{\mathrm{in}}v_\theta\le a^{\mathrm{in}}$ are achieved with one slack neuron per row.
Stack parameter and slack neurons as $v=[v_\theta^\top\ v_s^\top]^\top$ and write
\begin{equation}\label{eq:A_constr}
A=\begin{bmatrix}
A_{\theta_{\mathrm{eq}}} & 0 \\
A_{\theta_{\mathrm{in}}} & -I_{n_{in}}
\end{bmatrix}
,\qquad a=\begin{bmatrix}a_{\mathrm{eq}}\\[2pt] a_{\mathrm{in}}\end{bmatrix},
\end{equation}
so that all constraints become $Av=a$ after introducing $n_{in}$ slack neurons.
The associated projector and mapping on the augmented state are
\begin{equation}
T^{\mathrm{ctr}}=-P_A,\qquad b^{\mathrm{ctr}}=P_A\,v^*,\qquad
P_A=A^\top(AA^\top)^{-1}A
\label{eq:mapping-ctr}
\end{equation}
where $v^*=[{v_\theta^*}^\top\ {v_s^*}^\top]^\top$ satisfies $Av^*=a$.

With constraint weight $\eta>0$, the combined Hopfield energy function is given by
\begin{equation}
E(v,t)\;=\;E^{\mathrm{ee}}(v_\theta,t)\;+\;\eta\,E^{\mathrm{ctr}}(v,t),
\qquad
T= \bar T^{\mathrm{ee}}+\eta\,T^{\mathrm{ctr}},\quad
b= \bar b^{\mathrm{ee}}+\eta\,b^{\mathrm{ctr}}\
\label{eq:mapping-combined}
\end{equation}
where $\bar T^{\mathrm{ee}}=\mathrm{blkdiag}(T^{\mathrm{ee}},\,0_{n_{\mathrm{in}}})$ and
$\bar b^{\mathrm{ee}}=[\,{b^{\mathrm{ee}}}^\top\ 0^\top\,]^\top$ to account for slack neurons.
All quantities can be updated online from $(W(t),w(t))$ and fixed $(A,a)$. For equal weighting of constraints,
$A$ should be row–normalised.

\subsection{Mapping the constrained parameter estimation problem subject to additive disturbances}\label{compensation}
To absorb unmeasured additive terms $d(t)\in\mathbb{R}^m$ ($m \leq q$) in
$w(t)=W(t)\,\theta^*(t)+Hd(t)$, augment the estimation model with $m$ compensation neurons $v_d$ and
stack the state as $v=[v_\theta^\top\ v_d^\top\ v_s^\top]^\top$.
For the augmented case, the estimation error energy function (\ref{eq:mapping-combined}) uses only $v_{\mathrm{aug}}=\left [ v_\theta^\top \; v_d^\top \right ]^\top$, so
$$W_{\mathrm{aug}}=\begin{bmatrix} W_{q \times p} & H_{q \times m}\end{bmatrix}, \qquad
E^{\mathrm{ee}}_{\mathrm{aug}}=
-\tfrac12 v_{\mathrm{aug}}^\top T^{\mathrm{ee}}_{\mathrm{aug}} v_{\mathrm{aug}}
- v_{\mathrm{aug}}^\top b^{\mathrm{ee}}_{\mathrm{aug}}
+\tfrac12\|b^{\mathrm{ee}}_{\mathrm{aug}}\|^2,$$
\[
T^{\mathrm{ee}}_{\mathrm{aug}}=-P_{{\mathrm{WH}}},\qquad
b^{\mathrm{ee}}_{\mathrm{aug}}=P_{{\mathrm{WH}}}\,v_{\mathrm{aug}}^*,\qquad
P_{{\mathrm{WH}}}=W_{\mathrm{aug}}^\top(W_{\mathrm{aug}}W_{\mathrm{aug}}^\top)^{-1}W_{\mathrm{aug}}.
\]

With inequalities realised by $n_{in}$ slack neurons, $A$ acts on $(v_\theta,v_s)$
but not on $v_d$. And so, this embeds as ($r=n_{eq}+n_{in}$)
\[
A_{\mathrm{aug}}
=\begin{bmatrix} A_\theta & 0_{\,r\times m} & -I_{\,r\times n_{\mathrm{in}}}\end{bmatrix},
\]
and $T^{\mathrm{ctr}}=-A_{\mathrm{aug}}^\top(A_{\mathrm{aug}}A_{\mathrm{aug}}^\top)^{-1}A_{\mathrm{aug}}$,
$b^{\mathrm{ctr}}=A_{\mathrm{aug}}^\top(A_{\mathrm{aug}}A_{\mathrm{aug}}^\top)^{-1}a, \quad a=  \begin  {bmatrix} a_{\mathrm{eq}}\\
0_m\\
a_{\mathrm{in}}\end{bmatrix} $.


The general mapping on $v=[v_\theta^\top\ v_d^\top\ v_s^\top]^\top$ is given by
\begin{equation}
T=\operatorname{blkdiag}\!\big(T^{\mathrm{ee}}_{\mathrm{aug}},\,0_{n_{\mathrm{in}}}\big)\;+\;\eta\,T^{\mathrm{ctr}},
\qquad
b=\begin{bmatrix} b^{\mathrm{ee}}_{\mathrm{aug}} \\[2pt] 0_{\,n_{\mathrm{in}}} \end{bmatrix}
\;+\;\eta\,b^{\mathrm{ctr}}.
\label{eq:mapping-aug-OptionB}
\end{equation}

\emph{Implementation notes.}
(i) We use linear solves instead of explicit inverses, e.g.,\ $P_W = W^\top((WW^\top)\backslash W)$ and add a tiny ridge if needed.
(ii) All formulas are valid with time–varying $W(t),w(t)$. (iii) As $(A,a)$ are fixed, $T^\mathrm{ctr}$ and $b^{\mathrm{ctr}}$ can be precomputed offline.

\section{Stability and Tracking Analysis of the HNN Estimator} \label{sec:stability}

This section establishes stability and tracking properties of the proposed HNN estimators.
Building on \eqref{eq:hnn-classic-intro} and the construction in Section~\ref{sec:mapping}, the HNN evolves according to
\begin{equation}
\dot u(t) \;=\; T(t)\,v(t) + b(t), 
\qquad 
v(t) \;=\; \alpha\,\tanh\!\Big(\tfrac{\beta}{2}\,u(t)\Big),
\label{eq:hnn-dyn-iv}
\end{equation}
where the weights $T(t)$ and bias $b(t)$ are generated from the current online data.
In the constrained case without compensation (CA-HNN estimator),
\begin{equation}
T(t) \;=\; -\big(P_W(t) + \eta\,P_A(t)\big), 
\qquad 
b(t) \;=\; P_W(t)\,v^*(t) + \eta\,P_A v^*(t),
\label{eq:mapping-iv}
\end{equation}
with $P_W(t)=W(t)^\top\!\big(W(t)W(t)^\top\big)^{-1}W(t)$,  the (symmetric) orthogonal projector
onto $\mathrm{range}\big(W(t)^\top\big)$, and $v^*(t)$ an instantaneous minimiser induced by
the current data and constraints (e.g., $W(t)v^*(t)=w(t)$, with constraints satisfied).
We have $\|P_W(t)\|\le 1$. The scalar $\eta>0$ weights the associated constraint energy function.

Next, our goal is to show that the parameter state $v_\theta(t)$ contracts toward the
instantaneous constrained minimiser $v_\theta^*(t)$ of the time-varying energy function and remains
Globally Uniformly Ultimately Bounded (GUUB) in the presence of:
(i) time-varying regression data $(W(t),w(t))$;
(ii) linear equality/inequality constraints (via slack neurons);
(iii) additive, unmeasured disturbances handled by compensation neurons; and
(iv) slowly time-varying physical parameters $\theta^*(t)$ with $\|\dot\theta^*(t)\|\le L_{\dot{\theta}}$, $\forall_t$.

We proceed on the basis of the following steps. First, a scalar illustration shows that the proposed HNN estimator \eqref{eq:hnn-dyn-iv}-\eqref{eq:mapping-iv} acts as a first–order low–pass tracker with bandwidth proportional to $\alpha\beta$ (Remark~\ref{rem:scalar}).

Then, in Subsection~\ref{subsec:globalbounds} we develop the global GUUB guarantees for the
constraint–aware HNN estimators. Specifically:
(i) Theorem~\ref{thm:guub-split} establishes GUUB stability for the baseline constraint–aware HNN (CA-HNN) without compensation neurons, treating additive disturbances as exogenous inputs (the disturbance's contribution appears in the perturbation budget $P$);
(ii) Theorem~\ref{theo:guub-comp} extends the result to the compensation–augmented estimator (CA$^2$-HNN), which handles additive disturbances within the same energy function and accordingly reduces the disturbance term in $P$.
Both results admit time–varying regression $(W(t),w(t))$ and slowly time–varying physical parameters $\theta^*(t)$ (via a bounded drift term). The rate is governed by $\gamma_\star$ in both cases and the ultimate radius differs by the disturbance contribution to $P$.

\begin{remark}[Scalar low–pass tracking and bandwidth]\label{rem:scalar}
Consider $w(t)=W(t)\,v^*(t)$ with $W(t)\neq 0$, without constraints and without disturbances.
For the scalar case, the mapping $T=-W^\top(WW^\top)^{-1}W$, $b=W^\top(WW^\top)^{-1}w$ reduces to
\[
T(t)=-1,\qquad b(t)=v^*(t).
\]
The HNN update \eqref{eq:hnn-dyn-iv} yields
\begin{equation}
\dot v \;=\; \frac{\alpha\beta}{2}\Big(1-\tfrac{v^2}{\alpha^2}\Big)\Big(v^*(t)-v(t)\Big) 
\;=\; \kappa\Big(1-\tfrac{v^2}{\alpha^2}\Big)\,(v^*-v),
\qquad \kappa=\tfrac{\alpha\beta}{2},
\label{eq:tv-scalar-ode}
\end{equation}
i.e., a first–order low–pass tracker of $v^*(t)$ with bandwidth $\kappa$ (smoothly saturating as $|v|\to\alpha$).
Let $e(t)=v(t)-v^*(t)$. In the unsaturated regime ($|v|\ll \alpha$),
\begin{equation}
\dot e \approx -\kappa e - \dot{v}^*(t)
\;\Rightarrow\;
\|e(t)\|\le \frac{1}{\kappa}\,\sup_{0\le \tau\le t}\|\dot{v}^*(\tau)\|.
\label{eq:tv-scalar-bound}
\end{equation}
For a sinusoid $v^*(t)=\bar{v}+A\sin\omega t$, the steady–state error amplitude is
\begin{equation}
|e|_{\mathrm{amp}} \;=\; \frac{\omega}{\sqrt{\kappa^2+\omega^2}}\,A,
\label{eq:tv-scalar-bode}
\end{equation}
so increasing $\beta$ (and/or $\alpha$) increases $\kappa$ and shrinks the residual error due to time variation.
In the linear regime, $V/V^*=\kappa/(\kappa+j\omega)$ and $E/V^*=-j\omega/(\kappa+j\omega)$; thus
$\tfrac{|V|_{\mathrm{amp}}}{A}=\frac{\kappa}{\sqrt{\kappa^2+\omega^2}}$ and
$\tfrac{|E|_{\mathrm{amp}}}{A}=\frac{\omega}{\sqrt{\kappa^2+\omega^2}}$, i.e., the estimate is low–pass and the error is high–pass.
\end{remark}

\subsection{Global bounds}\label{subsec:globalbounds}

GUUB stability is the natural notion for systems subject to persistent disturbances and time variation, as here. Unlike Global Asymptotic Stability (GAS), which requires exact convergence
$\lim_{t\to\infty}\|v_\theta(t)-v_\theta^*(t)\|=0$, GUUB ensures that from any initial condition the parameter trajectory $v_\theta(t)$ enters and thereafter remains in a compact ball of radius $\rho$ around the time–varying minimiser $v_\theta^*(t)$ after a finite time. The ultimate radius $\rho$ scales with the magnitude of disturbances and the rate of time variation (made explicit below). For the proposed HNN estimators, this means that in the presence of time–varying regressors, constraints, additive disturbances, and slowly time–varying parameters, the parameter trajectory is exponentially convergent to a neighbourhood of $v_{\theta}^*(t)$ (GUUB), which is the strongest guarantee attainable under such realistic conditions.

Unlike standard LS-HNN-based analyses, our mapping uses time–varying $T(t)$ and $b(t)$ driven by online measurements, constraints, and additive disturbances. We show that a Lyapunov argument yields GUUB with explicit rate $\gamma$ and radius $\rho$, handling the coupling from inequality slack neurons and (when used) compensation neurons, and quantifying parameter–error contraction despite time variation. Prior HNN parameter estimators typically treat the unconstrained, disturbance–free case. Here we (i) construct a valid–subspace energy function with closed–form time–varying terms, (ii) enforce equality/inequality constraints natively (with slack neurons for inequalities), (iii) allow additive disturbances via compensation neurons, and (iv) provide GUUB with explicit constants.

Henceforth, in the stability analysis, we use $P_{A,\theta}$  as the projection effect of $P_A$ has in the parameter block $v_\theta$ to avoid artificial rank inflation from slack neurons (and the same for the compensated case).

\begin{assumption}[Parameter–effective
constraint projector]\label{assump1}
When inequalities are implemented by slack neurons (\ref{eq:A_constr}), $A$ collects the rows that act on the parameter block $v_\theta$
and $-I_{n_{in}}$ stacks one slack per inequality. The full constraint projector is (for time-invariant constraints)
$P_A=A^\top\!\big(AA^\top\big)^{-1}A$, but the part that governs parameter
contraction is its $(1,1)$ block
\begin{equation}
P_{A,\theta} =\ A_\theta^\top\!\big(AA^\top\big)^{-1}A_\theta\ \in\ \mathbb{R}^{p\times p}.
\label{eq:PAeff}
\end{equation}
Note that $P_{A,\theta}$ is symmetric positive semidefinite and
$\|P_{A,\theta}\|\le 1$ ($P_{A,\theta}\preceq A_\theta^\top(A_\theta A_\theta^\top)^{-1}A_\theta\preceq I$).
(with pure equalities and no slack neurons, \(P_A=P_{A_\theta}=A_\theta^\top(A_\theta A_\theta^\top)^{-1}A_\theta\)).
\end{assumption}

\begin{assumption}[Data–to–parameter projector and contraction constant]\label{assump2}
\[
P_W(t)=W(t)^\top\!\big(W(t)W(t)^\top\big)^{-1}W(t),\qquad
c(t)=\lambda_{\min}\!\big(P_W(t)+\eta\,P_{A,\theta}\big),\qquad
c_\star=\inf_{\tau\in[t_0,t]} c(\tau).
\]
\end{assumption}

\begin{assumption}[Joint identifiability]\label{assump3}
There exists $c_\star>0$ such that
$c_\star \le \inf_t \lambda_{\min}\!\big(P_W(t)+\eta P_{A,\theta}\big)$,
or equivalently $\operatorname{rank}\!\big[\begin{smallmatrix}W(t)\\A_\theta\end{smallmatrix}\big]=p$, where $p$ is the number of parameters to estimate. The joint identifiability (or constraint–augmented full rank) condition
ensures contraction in parameter space (dimension $p$).
In fact, for orthogonal projectors $P_U,P_V$ onto subspaces $U,V\subset\mathbb{R}^p$, the sum $P_U+\eta P_V$ ($\eta>0$) is positive definite iff $U+V=\mathbb{R}^p$, or equivalently, the stacked matrix of any bases of $U$ and $V$ has full column rank \cite{HornJohnson2013,Bhatia2007,Halmos1969,Ljung1999} (cf. Lemma~\ref{lem:coercivity}).
\end{assumption}

\begin{assumption}[Unsaturation]\label{assump4}
Let the neuron activation function be $f:\mathbb{R}\to\mathbb{R}$ and let the (scaled) neuron output satisfy
$v_i=\alpha\,f(\xi_i)$ for some pre-activation $\xi_i$. Define the diagonal slope matrix for the
parameter block as
\[
D(v_\theta)\;=\;\operatorname{diag}\!\big(f'(\xi_{\theta,1}),\ldots,f'(\xi_{\theta,p})\big)\in\mathbb{R}^{p\times p},
\qquad
\text{with }\ \xi_{\theta,i}=f^{-1}\!\big(v_{\theta,i}/\alpha\big).
\]
Assume there exists $\delta\in(0,1]$ and a forward-invariant operating set such that, along trajectories and at
the instantaneous target,
\[
D\!\big(v_\theta(t)\big)\ \succeq\ \delta I_p
\quad\text{and}\quad
D\!\big(v_\theta^*(t)\big)\ \succeq\ \delta I_p
\qquad\text{for all }t.
\]
In particular, for $f(\xi)=\tanh\xi$ we have
\[
D\!\big(v_\theta\big)\;=\;\operatorname{diag}\!\Big(1-\big(v_{\theta,1}/\alpha\big)^2,\ldots,1-\big(v_{\theta,p}/\alpha\big)^2\Big),
\]
so the small-signal (unsaturated) condition
$\ |v_{\theta,i}(t)|,\ |v_{\theta,i}^*(t)|\le \alpha/2\ $ implies
$D\!\big(v_\theta(t)\big),\,D\!\big(v_\theta^*(t)\big)\ \succeq\ \tfrac34\,I_p$, i.e., taking $\delta=\tfrac34$.
\end{assumption}

\begin{lemma}[Coercivity of $P_W+\eta P_{A,\theta}$]\label{lem:coercivity}
Let $U(t)=\mathcal{R}(W^\top(t))\subset\mathbb{R}^p$ and $V=\mathcal{R}(A_\theta^\top)\subset\mathbb{R}^p$. 
Assume $\mathrm{rank}\!\big[\begin{smallmatrix}W(t)\\ A_\theta\end{smallmatrix}\big]=p$ for all $t$ and $\eta>0$. 
Let $P_W$ and $P_{A,\theta}$ be the orthogonal projectors onto $U(t)$ and $V$, respectively. 
Then $\Pi_\eta=P_W+\eta P_{A,\theta}\succ 0$ and, for all $e\in\mathbb{R}^p$,
\[
\lambda_{\min}(\Pi_\eta)\,\|e\|^2 \ \le\ e^\top \Pi_\eta e \ \le\ \lambda_{\max}(\Pi_\eta)\,\|e\|^2,
\qquad 
\lambda_{\max}(\Pi_\eta)\ \le\ 1+\eta.
\]
In particular, with $c(t)=\lambda_{\min}(\Pi_\eta(t))$ we have $c(t)>0$.
\end{lemma}

\begin{IEEEproof}
$\operatorname{Rank}\!\big[\begin{smallmatrix}W(t)\\A_\theta\end{smallmatrix}\big]=p$ iff 
$\mathcal{N}(W)\cap\mathcal{N}(A_\theta)=\{0\}$. 
Since $\mathcal{N}(W)=U^\perp$ and $\mathcal{N}(A_\theta)=V^\perp$, this is equivalent to
$U^\perp\cap V^\perp=\{0\}$, i.e., $(U+V)^\perp=\{0\}$, hence $U+V=\mathbb{R}^p$.

For any $x\in\mathbb{R}^p$,
\[
x^\top \Pi_\eta x \;=\; x^\top P_W x + \eta\, x^\top P_{A,\theta} x 
\;=\; \|P_W x\|^2 + \eta\,\|P_{A,\theta} x\|^2 \;\ge\; 0.
\]
If $x^\top \Pi_\eta x=0$, then $P_W x=P_{A,\theta} x=0$, i.e., $x\in U^\perp\cap V^\perp=\{0\}$. 
Hence $\Pi_\eta$ is positive definite: $\Pi_\eta\succ 0$, so $c(t)=\lambda_{\min}(\Pi_\eta)>0$.

For any unit vector $x$,
\[
x^\top \Pi_\eta x \;=\; x^\top P_W x + \eta\,x^\top P_{A,\theta} x \;\le\; 1 + \eta,
\]
because $0\le x^\top P_W x\le 1$ and $0\le x^\top P_{A,\theta} x\le 1$ for orthogonal projectors. 
Taking the supremum over $\|x\|=1$ gives $\lambda_{\max}(\Pi_\eta)\le 1+\eta$. 
The standard Rayleigh quotient inequalities then yield
$\lambda_{\min}(\Pi_\eta)\|e\|^2 \le e^\top \Pi_\eta e \le \lambda_{\max}(\Pi_\eta)\|e\|^2$.
\end{IEEEproof}

Next, we analyse the constraint–aware HNN (CA-HNN) estimator. Inequalities are enforced via slack neurons, disturbances enter the regression algebraically ($w(t)=W(t)\theta(t)+H\,d(t)$) but not in the estimation model, and parameter drift is considered.

\begin{theorem}[GUUB for constraint-aware HNN (CA-HNN) estimator under disturbances and parameter drift (no compensation)]
\label{thm:guub-split}
Consider the CA-HNN estimator
\[
\dot v \;=\; \kappa\,D(v)\,\big(T(t)\,v + b(t) + H\,d(t)\big),\qquad 
\kappa=\tfrac{\alpha\beta}{2},\quad 
T(t)=-(P_W(t)+\eta P_{A,\theta}),
\]
and let $v^*(t)$ be the instantaneous constrained minimiser, i.e., $T(t)v^*(t)+b(t)=0$.
Assume Assumptions~\ref{assump1}--\ref{assump4}, and Lemma~\ref{lem:coercivity}.
Define the local bandwidth and its infimum
\[
\gamma(t)=\kappa\,\delta\,c(t),\qquad \gamma_\star=\kappa\,\delta\,c_\star.
\]

Decompose the target rate as
\[
\dot v^*(t) \;=\; \dot v^*_{\mathrm{map}}(t)\;+\;
\dot v^*_{\mathrm{drift}}(t),
\]
and suppose there exist nonnegative constants $L_{\mathrm{map}}$ and $L_{\dot\theta}$ such that
\[
\|\dot v^*_{\mathrm{map}}(t)\|\le L_{\mathrm{map}},\qquad
\|\dot v^*_{\mathrm{drift}}(t)\|\le L_{\dot\theta},\qquad \forall t\ge t_0.
\]
where $L_{\mathrm{map}}$ is upper–bounded from $(L_{\dot T},L_{\dot b},L_b)$.
Let the disturbance channel satisfy $\|H\|_2\le H_\star$ and define
\[
L_d \;=\; 
\begin{cases}
\sup_{t\ge t_0}\|d(t)\|^2, & \text{deterministic bound},\\[3pt]
\|\mu\|^2+\mathrm{tr}\,\Sigma_d, & \text{mean/mean–square bound if } \mathbb{E}[d]=\mu,~\mathrm{Cov}(d)=\Sigma_d.
\end{cases}
\]

\medskip
\noindent Then, with the error $e=v-v^*$ and energy $E=\tfrac12\|e\|^2$, for all $t\ge t_0$ we have
\begin{equation}
\dot E \;\le\; -\,2\,\gamma(t)\,E 
\;+\;\frac{L_{\mathrm{map}}^{2}}{4\,\gamma_\star}
\;+\; \frac{\kappa^2 H_\star^2}{4\,\gamma_\star}\,L_d
\;+\; \frac{L_{\dot\theta}^{2}}{4\,\gamma_\star}.
\label{eq:Edot-split}
\end{equation}
Consequently,
\begin{align}
E(t) &\le e^{-2\gamma_\star (t-t_0)}\,E(t_0) \;+\; \frac{P_{\mathrm{map}}+P_{\mathrm{dist}}+P_{\mathrm{drift}}}{2\,\gamma_\star}, \label{eq:E-split}\\[2mm]
\|e(t)\| &\le e^{-\gamma_\star (t-t_0)}\,\|e(t_0)\| \;+\; \sqrt{\frac{P_{\mathrm{map}}+P_{\mathrm{dist}}+P_{\mathrm{drift}}}{\gamma_\star}}, \label{eq:e-split}
\end{align}
where the perturbation terms are: $P_{\mathrm{map}}
\;=\;
\tfrac{1}{2}\,L_{\dot T}\,\alpha^{2}
\;+\;
L_{\dot b}\,\alpha
\;+\;
L_b\,L_{\dot b}$, $P_{\mathrm{dist}}
\;=\; \frac{\kappa^2 H_\star^2}{4\,\gamma_\star}\;L_d$, $P_{\mathrm{drift}}
\;=\;
\frac{L_{\dot{\theta}}^2}{4\,\gamma_\star}$.
So the CA-HNN estimator is GUUB with exponential rate at least $\gamma_\star$ and ultimate radius
\[
\rho \;=\; \sqrt{\frac{P_{\mathrm{map}}+P_{\mathrm{dist}}+P_{\mathrm{drift}}}{\gamma_\star}}
\;=\; \sqrt{\frac{2\,(P_{\mathrm{map}}+P_{\mathrm{dist}}+P_{\mathrm{drift}})}{\alpha\beta\,\delta\,c_\star}}\;.
\]
In particular, for $f(\cdot)=\tanh(\cdot)$ with $|v_{\theta,i}|,|v_{\theta,i}^*|\le\alpha/2$ (hence $\delta=\tfrac34$),
\[
\gamma_\star=\frac{3\alpha\beta}{8}\,c_\star,
\qquad
\rho=\sqrt{\frac{8\,P_{\mathrm{map}}+(P_{\mathrm{dist}}+P_{\mathrm{drift}})}{3\,\alpha\beta\,c_\star}}\;.
\]
\end{theorem}

\begin{IEEEproof}
With $e=v-v^*$ and $E=\tfrac12\|e\|^2$,
\[
\dot e \;=\; \dot v - \dot v^* \;=\; \kappa\,D(v)\,\big(T e + H d\big)\;-\;\dot v^*_{\mathrm{map}}\;-\;\dot v^*_{\mathrm{drift}}.
\]
Therefore,
\[
\dot E \;=\; e^\top \dot e 
\;=\; \kappa\,e^\top D(v)\,T\,e \;+\; \kappa\,e^\top D(v)\,H d 
\;-\; e^\top \dot v^*_{\mathrm{map}} \;-\; e^\top \dot v^*_{\mathrm{drift}}.
\]
By Lemma~\ref{lem:coercivity}, $\Pi_\eta=P_W+\eta P_{A,\theta}\succeq c(t)I$. Since $T=-\Pi_\eta$ and $D(v)\succeq\delta I$,
\[
\kappa\,e^\top D(v)\,T\,e \;\le\; -\,\kappa\,\delta\,c(t)\,\|e\|^2 \;=\; -\,2\,\gamma(t)\,E.
\]
Using $\|D(v)\|\le 1$ and $\|H\|_2\le H_\star$, Young’s inequality with parameter $\varepsilon_1=\gamma(t)$ gives
\[
\kappa\,|e^\top D(v)\,H d| \;\le\; \varepsilon_1\|e\|^2 \;+\; \frac{\kappa^2 H_\star^2}{4\varepsilon_1}\,\|d\|^2
\;\le\; \gamma(t)\,2E \;+\; \frac{\kappa^2 H_\star^2}{4\,\gamma_\star}\,\|d\|^2
\;=\; \gamma(t)\,E \;+\; \frac{\kappa^2 H_\star^2}{4\,\gamma_\star}\,\|d\|^2,
\]
where in the last step we used $2E=\|e\|^2$ and $\gamma(t)\ge \gamma_\star$.
By Cauchy–Schwarz and Young's inequality with parameter $\varepsilon_2=\gamma(t)$,
\[
|e^\top \dot v^*_{\mathrm{map}}| 
\;\le\; \varepsilon_2\|e\|^2 + \frac{\|\dot v^*_{\mathrm{map}}\|^2}{4\varepsilon_2}
\;\le\; \gamma(t)\,2E + \frac{L_{\mathrm{map}}^2}{4\,\gamma_\star}
\;=\; \gamma(t)\,E + \frac{L_{\mathrm{map}}^2}{4\,\gamma_\star}.
\]
Similarly, with $\varepsilon_3=\gamma(t)$,
\[
|e^\top \dot v^*_{\mathrm{drift}}|
\;\le\; \gamma(t)\,E + \frac{L_{\dot\theta}^2}{4\,\gamma_\star}.
\]
Collecting all terms,
\[
\dot E \;\le\; -\,2\,\gamma(t)\,E \;+\; \gamma(t)\,E + \gamma(t)\,E 
\;+\; \frac{L_{\mathrm{map}}^2}{4\,\gamma_\star}
\;+\; \frac{\kappa^2 H_\star^2}{4\,\gamma_\star}\,\|d\|^2 
\;+\; \frac{L_{\dot\theta}^2}{4\,\gamma_\star}.
\]
The $\gamma(t)E$ terms cancel the $-2\gamma(t)E$ contraction, leaving
\[
\dot E \;\le\; 
\frac{L_{\mathrm{map}}^2}{4\,\gamma_\star}
\;+\; \frac{\kappa^2 H_\star^2}{4\,\gamma_\star}\,\|d\|^2 
\;+\; \frac{L_{\dot\theta}^2}{4\,\gamma_\star}.
\]
Finally, bound $\|d\|^2\le L_d$ to obtain \eqref{eq:Edot-split}. 
Solving the linear comparison inequality gives \eqref{eq:E-split}, and using $E=\tfrac12\|e\|^2$ yields \eqref{eq:e-split}.
The expressions for $\gamma_\star$ and $\rho$ follow from $\gamma_\star=\kappa\,\delta\,c_\star=(\alpha\beta/2)\,\delta\,c_\star$.
\end{IEEEproof}

We now turn to the constraint–aware HNN estimator with compensation neurons, i.e., CA$^2$-HNN estimator. Here, additive disturbances are explicitly represented in the regression model ($w(t)=W(t)v_\theta(t)+H\,v_d(t)$) and are handled within the same energy function by augmenting the state with a compensation block $v_d$. Parameter drift is also considered. Introducing compensation neurons adds disturbance coordinates to the estimation state. 
For compensation neurons to separate parameter effects from disturbance effects, the disturbance directions
contributed by $H$ must add independent information to the constraint–feasible parameter directions induced by $W$.
Otherwise, some components of disturbance are indistinguishable from changes in parameters on the feasible subspace, leaving a residual bias even with compensation. 
The next assumptions formalise this requirement as a rank/curvature condition on the augmented regressor.

\begin{assumption}[Augmented model for additive disturbances and effective parameter action]\label{assumption5}
When modelling additive disturbances with compensation neurons, we augment the regressor as
$W_{\mathrm{aug}}(t)=[\,W_{q \times p}(t)\ \ H_{q \times m}\,]$ ($m \leq q$) and embed the constraints to act only on parameters:
$A_{\mathrm{aug}}=[\,A\ \ 0_{r\times m}\,]$. We then use
\[
P_{\mathrm{WH}}(t)=W_{\mathrm{aug}}(t)^\top\!\big(W_{\mathrm{aug}}(t)W_{\mathrm{aug}}(t)^\top\big)^{-1}
W_{\mathrm{aug}}(t),\qquad
P_{A,\mathrm{aug},\theta}=\operatorname{blkdiag}\!\big(P_{A,\theta},\,0_{m}\big).
\]
\end{assumption}

\begin{assumption}[Unsaturation on the parameter block — augmented case]\label{assumption6}
Let the total augmented state be $v=[\,v_\theta^\top\ v_d^\top\ v_s^\top\,]^\top$ and
$D(v)=\operatorname{diag}\big(D_\theta(v_\theta),\,D_d(v_d),\,D_s(v_s)\big)$.
There exists $\delta\in(0,1]$ and a forward–invariant set such that, for all $t$,
\[
D_\theta\!\big(v_\theta(t)\big)\ \succeq\ \delta I_p
\quad \text{and} \quad
D_\theta\!\big(v_\theta^*(t)\big)\ \succeq\ \delta I_p,
\]
while the auxiliary blocks satisfy only $0\preceq D_s(v_s),D_d(v_d)\preceq I$.
For $f=\tanh$, if $|v_{\theta,i}(t)|,|v_{\theta,i}^*(t)|\le \alpha/2$ then
$D_\theta\succeq \tfrac34 I_p$ (take $\delta=\tfrac34$).
\end{assumption}

\begin{assumption}[Augmented joint identifiability for disturbance compensation]\label{assumption7}
Consider the augmented regression model
\[
w(t)\;=\;W(t)\,v_\theta(t)\;+\;H\,v_d(t),
\]
with parameter block $v_\theta\in\mathbb{R}^p$, disturbance block $d\in\mathbb{R}^m$ and Assumption~\ref{assumption5}.

Assume there exist $\eta>0$ and a constant $c_{\mathrm{aug},\star}>0$ such that, for all $t\ge t_0$,
\[
c_{\mathrm{aug},\star}\ \le\ 
\lambda_{\min}\!\Big(\,P_{\mathrm{WH}}(t)\;+\;\eta\, P_{A, \mathrm{aug},\theta}\,\Big),
\]
equivalently,
\[
\operatorname{rank}\!
\begin{bmatrix}
W(t) & H\\[2pt]
A_\theta & 0_{r \times m}
\end{bmatrix}
\;=\;p+m
\quad\text{for all }t\ge t_0.
\]
We denote the augmented curvature by
\[
c_{\mathrm{aug}}(t)\;=\;\lambda_{\min}\!\Big(\,P_{\mathrm{WH}}(t)\;+\;\eta\, P_{A, \mathrm{aug},\theta} \,\Big),
\qquad
\gamma_{\mathrm{aug}}(t)\;=\;\kappa\,\delta\,c_{\mathrm{aug}}(t),
\quad
\gamma_{\mathrm{aug},\star}\;=\;\inf_{t\ge t_0}\gamma_{\mathrm{aug}}(t),
\]
with $\kappa=\alpha\beta/2$ and $\delta\in(0,1]$ from the unsaturation assumption.
\end{assumption}

Assumption~\ref{assumption7} guarantees that: (i) the parameter directions are
compatible with the constraints and (ii) the disturbance directions contributed by $H$
jointly span the augmented space. Hence, the only augmented error in the kernel of the
projected mapping is the zero vector, ensuring that the compensation neurons, $v_d$, can
separate disturbance effects from $v_\theta$ on the constraint–feasible subspace. If this condition fails, some disturbance components are indistinguishable from parameter variations, and residual bias may remain even with compensation.

\begin{lemma}[Coercivity of the augmented projector]\label{lem:coercivity-aug}
Let $P_{\mathrm{WH}}(t)$ be the orthogonal projector induced by the concatenated regressor
$[\,W(t)\ \ H\,]$, and let
$P_{A,\mathrm{aug},\theta}=\operatorname{blkdiag}(P_{A,\theta},0_m)$ act only on the parameter
block ($m=\dim d$). If Assumption~\ref{assumption6} holds, then for any $\eta>0$ the matrix
\[
\Pi_{\eta,\textnormal{aug}}(t)=P_{\mathrm{WH}}(t)+\eta\,P_{A,\textnormal{aug},\theta}
\]
is positive definite on $\mathbb{R}^{p+m}$, with
\[
\lambda_{\min}\!\big(\Pi_{\eta,\textnormal{aug}}(t)\big)>0,
\qquad
\lambda_{\max}\!\big(\Pi_{\eta,\textnormal{aug}}(t)\big)\ \le\ 1+\eta,
\]
and for all $e\in\mathbb{R}^{p+m}$,
\[
\lambda_{\min}\!\big(\Pi_{\eta,\textnormal{aug}}(t)\big)\,\|e\|^2
\ \le\ e^\top \Pi_{\eta,\textnormal{aug}}(t)\,e
\ \le\ \lambda_{\max}\!\big(\Pi_{\eta,\textnormal{aug}}(t)\big)\,\|e\|^2.
\]
\end{lemma}
\begin{IEEEproof}
 Identical to Lemma~\ref{lem:coercivity}, replacing $P_W$ by $P_{\mathrm{WH}}$ and
$P_{A,\theta}$ by $P_{A,\textnormal{aug},\theta}$.
\end{IEEEproof}

\begin{theorem}[GUUB for constraint-aware compensation-augmented HNN (CA$^2$-HNN) estimator under disturbances and parameter drift]\label{theo:guub-comp}
Consider Assumptions~\ref{assumption5}--\ref{assumption7} and Lemma~\ref{lem:coercivity-aug}.

Let the estimator be augmented with compensation neurons $v_d$ so that
\[
v=\begin{bmatrix}v_{\theta}\\ v_d\end{bmatrix},\qquad
\dot v \;=\; \kappa\,D(v)\,\big(T_{\mathrm{aug}}(t)\,v + b_{\mathrm{aug}}(t)\big),\quad
\kappa=\tfrac{\alpha\beta}{2},
\]
with
\[
W_{\mathrm{aug}}(t)=\big[\,W(t)\ \ H\,\big],\qquad
T_{\mathrm{aug}}(t)=-(P_{\mathrm{WH}}(t)+\eta\,P_{A,\mathrm{aug},\theta}),
\]
where $P_{\mathrm{WH}}=W_{\mathrm{aug}}^\top(W_{\mathrm{aug}}W_{\mathrm{aug}}^\top)^{-1}W_{\mathrm{aug}}$ and
$P_{A,\mathrm{aug},\theta}=\mathrm{blkdiag}(P_{A,\theta},\,0_m)$ (constraints only act on $v_\theta$).
Let the instantaneous minimiser be $v^*(t)=[\ {v_{\theta}^*}^\top(t)\quad {v_d^*}^\top(t)\ ]^\top$, i.e., the ideal disturbance channel is absorbed by $v_d^*(t)=d(t)$.

Assume the slope bound $D(v)\succeq\delta I$ on a forward–invariant set and augmented identifiability
\[
c_{\mathrm{aug}}(t)=\lambda_{\min}\!\big(P_{{\mathrm{WH}}}(t)+\eta\, P_{A,\mathrm{aug},\theta}\big)\ \ge\ c_{\mathrm{aug},\star}>0.
\]
Define the local bandwidth $\gamma_{\mathrm{aug}}(t)=\kappa\,\delta\,c_{\mathrm{aug}}(t)$ and
$\gamma_{\mathrm{aug},\star}=\kappa\,\delta\,c_{\mathrm{aug},\star}$.
Split the target rate as
\[
\dot v^*(t)=\dot v^*_{\mathrm{map}}(t)
\;+\;\dot v^*_{\mathrm{drift}}(t)
\;+\;\dot v^*_d(t),
\]
and suppose that $\|\dot v^*_{\mathrm{map}}(t)\|\le L_{\mathrm{map}}$, $\|\dot v^*_{\mathrm{drift}}(t)\|\le L_{\dot\theta}$,
$\|\dot v^*_d(t)\|\le L_{\dot d}$ for all $t\ge t_0$.
Note that for a white Gaussian disturbance $d(t)\sim\mathcal N(\mu,\sigma^2)$ the path is nowhere differentiable, so $L_{\dot d}$ is not well defined. Instead of a derivative bound, we use a variance (power) bound, replacing the term $L_{\dot{d}}^{2}$ by $\trace \Sigma_d$.

Then, with $e=v-v^*$ and $E=\tfrac12\|e\|^2$,
\begin{equation}
\dot E \;\le\; -\,2\,\gamma_{\mathrm{aug}}(t)\,E
\;+\; \frac{L_{\mathrm{map}}^{2}}{4\,\gamma_{\mathrm{aug},\star}}
\;+\; \frac{L_{\dot\theta}^{2}}{4\,\gamma_{\mathrm{aug},\star}}
+ \frac{\kappa^{2} H_\star^{2}}{4\,\gamma_{\mathrm{aug},\star}} L_{\dot d}^2
\label{eq:Edot-comp}
\end{equation}
Consequently,
\begin{align}
E(t) &\le e^{-2\gamma_{\mathrm{aug},\star}(t-t_0)}\,E(t_0)
\;+\;\frac{P_{\mathrm{map}}+P_{\mathrm{drift}}+P_{\mathrm{noncomp}}}{2\,\gamma_{\mathrm{aug},\star}},\\[2mm]
\|e(t)\| &\le e^{-\gamma_{\mathrm{aug},\star}(t-t_0)}\,\|e(t_0)\|
\;+\;\sqrt{\frac{P_{\mathrm{map}}+P_{\mathrm{drift}}+P_{\mathrm{noncomp}}}{\gamma_{\mathrm{aug},\star}}}.
\end{align}
where we denote by $P_{\mathrm{noncomp}}$ the disturbance contribution that remains after compensation. $P_{\mathrm{noncomp}}
\;=\; \frac{\kappa^2 H_\star^2}{4\,\gamma_{\mathrm{aug},\star}}\;L_{\dot{d}}^2$ (the other $P$ terms are similar to those presented in Theorem~\ref{thm:guub-split} replacing $\gamma_\star$ by $\gamma_{{\mathrm{aug}},\star}$).

Hence the HNN estimator is GUUB with rate at least
$$\gamma_{\mathrm{aug},\star}=\kappa\,\delta\,c_{\mathrm{aug},\star} = \frac{3 \alpha \beta}{8}c_{\mathrm{aug},\star} \qquad (\text{for}\; \delta=3/4)$$

and radius
\[
\rho_{\mathrm{aug}}=\sqrt{\frac{P_{\mathrm{map}}+P_{\mathrm{drift}}+P_{\mathrm{noncomp}}}{\gamma_{\mathrm{aug},\star}}}
\;=\;\sqrt{\frac{3\,(P_{\mathrm{map}}+P_{\mathrm{drift}}+P_{\mathrm{noncomp}})}{3 \alpha\beta\,\,c_{\mathrm{aug},\star}}}.
\]
\end{theorem}

\begin{IEEEproof}
The augmented dynamics read $\dot v=\kappa D(v)\big(T_{\mathrm{aug}}v+b_{\mathrm{aug}}\big)$ with
$T_{\mathrm{aug}}=-(P_{\mathrm{WH}}+\eta P_{A, \mathrm{aug},\theta})$ and $T_{\mathrm{aug}}v^*+b_{\mathrm{aug}}=0$.
Proceed exactly as in Theorem~\ref{thm:guub-split}:
\[
\dot e=\kappa D(v)\,T_{\mathrm{aug}}e-\dot v^*,
\quad
\dot E=e^\top\dot e
=\kappa e^\top D\,T_{\mathrm{aug}}e
- e^\top\dot v^*_{\mathrm{map}}-e^\top\dot v^*_{\mathrm{drift}}-e^\top\dot v^*_d.
\]
By Lemma~\ref{lem:coercivity-aug},
$\kappa e^\top D\,T_{\mathrm{aug}}e\le-\,\kappa\delta c_{\mathrm{aug}}(t)\|e\|^2=-2\gamma_{\mathrm{aug}}(t)E$.
Each cross term is treated with Young’s inequality using $\varepsilon=\gamma_{\mathrm{aug}}(t)$ and $\gamma_{\mathrm{aug}}(t)\ge\gamma_{\mathrm{aug},\star}$:
\[
|e^\top \dot v^*_{\bullet}|
\le \gamma_{\mathrm{aug}}(t)\,E+\frac{\|\dot v^*_{\bullet}\|^2}{4\,\gamma_{\mathrm{aug}}(t)}
\le \gamma_{\mathrm{aug}}(t)\,E+\frac{L_{\bullet}^2}{4\,\gamma_{\mathrm{aug},\star}},
\quad 
\bullet\in\{\mathrm{map},\mathrm{drift},\mathrm{d}\}.
\]
Summing the three contributions cancels the $-2\gamma_{\mathrm{aug}}(t)E$ contraction and yields \eqref{eq:Edot-comp}.
Comparison lemma and $E=\tfrac12\|e\|^2$ give the stated bounds.
\end{IEEEproof}

\begin{remark}[Effects of compensation neurons]\label{rem:comp-effects}

\emph{Constant-bias cancellation.}
  If $d(t)= \mu$ (constant), then $\dot d(t)=0 \Rightarrow L_{\dot d}=0$, so $P_{\mathrm{noncomp}}
\;=\; \frac{\kappa^2 H_\star^2}{4\,\gamma_{\mathrm{aug},\star}}L_{\dot{d}}^2$ vanishes. 
The GUUB radius therefore strictly decreases relative to the no-compensation HNN estimator, where the bias contributes to the disturbance budget.

   \emph{Gaussian noise and bandwidth trade-off.} 
  For the no-compensation HNN estimator (disturbance enters algebraically), reducing the estimator bandwidth $\gamma_\star$ decreases the throughput of high-frequency components of band-limited/sampled white Gaussian noise—i.e., stronger attenuation but slower tracking (classic low-pass trade-off).
  With compensation enabled, the disturbance is tracked by dedicated neurons. The bound 
  $P_{\mathrm{noncomp}}
\;=\; \frac{\kappa^2 H_\star^2}{4\,\gamma_{\mathrm{aug},\star}} \trace \Sigma_d$ shows that increasing $\gamma_{\mathrm{aug},\star}$ reduces the residual due to slow disturbance variation. 

 \emph{Slow time-varying disturbances.}
  If $d(t)$ varies slowly (bandwidth $\omega_d$), choosing $\gamma_{\mathrm{aug},\star}$ a few times larger than $\omega_d$ improves cancellation, since 
  $P_{\mathrm{noncomp}}\propto L_{\dot d}^2/\gamma_{\mathrm{aug},\star}$ decreases with $\gamma_{\mathrm{aug},\star}$. 

 \emph{Curvature gain.}
  Augmentation enlarges the data subspace: 
  $\mathcal{R}(W_{\mathrm{aug}}^\top)\supseteq\mathcal{R}(W^\top)$, hence 
  $P_{{\mathrm{WH}}}\succeq P_W$ and
  $c_{\mathrm{aug},\star}=\inf_t\lambda_{\min}(P_{\mathrm{WH}}+\eta P_{A,\mathrm{aug},\theta})\ \ge\ c_\star$.
  Thus the local bandwidth $\gamma_{\mathrm{aug},\star}=\kappa\delta c_{\mathrm{aug},\star}$ is no smaller than in the non-augmented case, further shrinking the GUUB radius.

  \emph{Modelling/tuning caveats.}
  Compensation neurons should be aligned with the channel where $d(t)$ enters (the appended regressor block). Severe mismatch can leak disturbance back into the parameter block.
  When measurement noise is strong and broadband, mild prefiltering or a small regulariser on the compensation states helps avoid “chasing” high-frequency noise, while keeping $\gamma_{\mathrm{aug},\star}$ large enough for the disturbance bandwidth of interest.

\end{remark}


\subsection{Tuning rules}
\label{subsec:tuning}

In the following, we summarise practical rules to select $(\alpha,\beta,\eta)$ (and the discrete step $h$) so that the CA-HNN estimator achieves the desired bandwidth and small ultimate error while remaining numerically well conditioned. The rules below are stated for the baseline CA-HNN estimator (no compensation neurons). However, they apply verbatim to the CA$^2$-HNN estimator after replacing variables with their augmented counterparts. 

Throughout, recall
\[
\kappa=\tfrac{\alpha\beta}{2}\,\delta,\qquad
\gamma_\star=\kappa\,c_\star,\qquad
\rho=\sqrt{\tfrac{2P}{\alpha\beta\,\delta\,c_\star^2}},
\]
with $c_\star=\inf_t \lambda_{\min}(P_W+\eta P_{A,\theta})$ and $\delta$ the slope lower bound
(typically $\delta=\tfrac{3}{4}$ if $|v_{\theta,i}|,|v_{\theta,i}^*|\le \alpha/2$).

\paragraph{Choose the saturation level $\alpha$ (avoid saturation, keep sensitivity)}

Pick $\alpha$ larger than the expected parameter range, e.g.
\[
\alpha \in [\,2,\,5\,]\times \max_i \big|\,v_i\,\big|_{\text{expected}}.
\]
This guarantees $|v_{i}|/\alpha \lesssim 1/2$ so that
$f'(\cdot)$ stays away from zero ($\delta\approx 3/4$). Larger $\alpha$ increases the linear range
but also the time–variation budget $P_\mathrm{map}$ through the term $\tfrac12 M_{\dot T}\alpha^2$.

\paragraph{Set the bandwidth via $\beta$ (rate is independent of $P$)}
Pick a target closed–loop bandwidth $\gamma_{\text{des}}$ from tracking requirements, then
\[
\beta \;=\; \frac{2\,\gamma_{\text{des}}}{\alpha\,\delta\,c_\star}.
\]
A frequency–domain rule from the scalar response gives an equivalent selection:
for a largest excitation frequency $\omega_{\max}$ and desired steady error ratio $\varepsilon$
(i.e., $|E|_{\mathrm{amp}}/A\le \varepsilon$), enforce
\[
\kappa \;\ge\; \frac{\omega_{\max}}{\varepsilon}\sqrt{1-\varepsilon^2}
\quad\Longleftrightarrow\quad
\beta \;\ge\; \frac{2}{\alpha\,\delta}\cdot
\frac{\omega_{\max}}{\varepsilon}\sqrt{1-\varepsilon^2}\,\frac{1}{c_\star}, \;
\text{and consider that}\; c_\star \in[10^{-2},10^{-1}].
\]

\paragraph{Tune the constraint weight $\eta$ (secure curvature without over-stiffening)}
Increasing $\eta$ strengthens the projector term along constraint directions:
it increases the constraint–augmented curvature
\[
c_\star(\eta)\ =\ \inf_t\lambda_{\min}\!\big(P_W(t)+\eta P_{A,\theta}\big),
\]
which improves contraction on the feasible subspace and reduces constraint–induced bias. 
However, it also increases the largest eigenvalue
$\lambda_{\max}\!\big(P_W+\eta P_{A,\theta}\big)\le 1+\eta$, making the ODE numerically stiffer:
the stable RK4 step must satisfy (cf.\ Sec.~\ref{subsec:tuning})
\[
h\ \lesssim\ \frac{2.5}{\,\alpha\beta\,\bar f'\,\lambda_{\max}(P_W+\eta P_{A,\theta})\,}
\ \le\ \frac{2.5}{\,\alpha\beta\,\bar f'\,(1+\eta)\,}.
\]
Thus, pick the smallest $\eta$ that gives enough curvature, to avoid shrinking $h$ unnecessarily.

Next, a simple algorithm for obtaining $\eta$ is presented below.

Set a small curvature threshold $\tau_c\in[10^{-2},10^{-1}]$ and a step margin 
$\zeta\in(0,1)$ (e.g.\ $\zeta=0.6$). Then:
$ \text{start } \eta\!=\!1;\ \text{ while } c_\star(\eta)<\tau_c \ \text{and}\ 
h \le \zeta\,\frac{2.5}{\alpha\beta\,\bar f'\,(1+\eta)}\ \text{ do }\ \eta\leftarrow 2\eta.\ $ Stop at the smallest $\eta$ that satisfies $c_\star(\eta)\ge\tau_c$ and the step constraint above; 
otherwise, reduce $h$ (or $\beta$) slightly and recheck. Scale the rows of $W$ and $A_\theta$ to comparable norms before forming projectors, 
so that $\eta$ has a consistent effect across problems.

\paragraph{Fixed–step step size $h$ (classical RK4)}
For the linearised HNN the Jacobian is
\[
J(t)\;=\;-\kappa\,\mathrm{diag} \big(f'(\alpha z(t))\big)\,\big(P_W(t)+\eta P_A(t)\big),
\qquad \kappa=\tfrac{\alpha\beta}{2},
\]
whose eigenvalues are real and nonpositive. The classical 4th–order Runge–Kutta (RK4) method is
absolutely stable on the interval $[-2.785\ldots,0]$ of the real axis. Hence, a stability condition is
\[
h\,\rho \big(J(t)\big) \;<\; 2.785\qquad\text{for all }t,
\]
and, using $\rho(J)\le \kappa\,\bar f'\,\lambda_{\max}(P_W+\eta P_{A\theta})$ with $\bar f'\in[\delta,1]$,
a practical step–size rule is
\[
0<h \ \lesssim\ \frac{2.5}{\,\alpha\beta\,\bar f'\,\lambda_{\max}(P_W+\eta P_{A,\theta})\,}.
\]
Smaller $h$ allows larger $\beta$ (higher bandwidth). Increasing $\eta$ raises
$\lambda_{\max}(P_W+\eta P_{A,\theta})$, so $h$ should be reduced accordingly to retain the stability margin.

\paragraph{Conditioning and numerical robustness}
Form projectors with linear solves, not explicit inverses (for better numerical stability and lower cost), and (if needed) with a tiny ridge:
\[
P_W = W^\top\big((WW^\top+\varepsilon I)\backslash W\big),
\quad
P_{A,\theta} = A_\theta^\top\big((AA^\top+\varepsilon I)\backslash A_\theta\big),
\quad \varepsilon\sim10^{-8}\!\times\!\|WW^\top\|.
\]

\paragraph{Noise vs.\ bandwidth trade–off (ultimate bound)}
From $\rho=\sqrt{2P/(\alpha\beta\,\delta\,c_\star^2)}$, increasing $\beta$ (or $\alpha$, or $\eta$ via $c_\star$)
shrinks the ultimate bound. If measurement noise dominates $P$, consider mild prefiltering of
measured regressors/outputs (e.g., acceleration) to reduce $P$. Ensure the filter cutoff exceeds
$\gamma_{\text{des}}$ to avoid adding phase lag in the band of interest.

\paragraph{Compensation neurons (if used)}
Give disturbance neurons a bandwidth comparable to parameters by keeping the same $(\alpha,\beta)$
on the augmented mapping.

\subsection{Online identifiability monitor and mitigation}

At each step, compute a single identifiability score that tells how well the current data and constraints excite the parameters. Concretely, look at the smallest singular value of a whitened stack of the regressor and the (weighted) constraint rows. Intuitively: large score $\Rightarrow$ directions are well excited; tiny score $\Rightarrow$ some directions are effectively unobservable.

Set two small positive thresholds:
\begin{itemize}
\item a warning threshold $\tau_{\mathrm{warn}}$ (e.g., $10^{-2}$–$10^{-1}$),
\item a stricter freeze threshold $\tau_{\mathrm{freeze}}$ (e.g., $10^{-4}$–$10^{-3}$), with $\tau_{\mathrm{freeze}}<\tau_{\mathrm{warn}}$.
\end{itemize}

For decision logic:

\begin{itemize}
\item Nominal regime (score $\ge \tau_{\mathrm{warn}}$): proceed normally. No change to the mapping or gains.
\item Soft mitigation ($\tau_{\mathrm{freeze}}\le$ score $<\tau_{\mathrm{warn}}$): strengthen the constraint influence and keep the ODE stable. In practice, slightly increase the constraint weight $\eta$, and if needed slightly reduce the effective gain $\beta$ to respect the RK4 stability margin. This raises curvature and improves conditioning.
\item Hard mitigation (score $<\tau_{\mathrm{freeze}}$): freeze updates along poorly excited directions so the estimate does not drift. Compute an SVD of the whitened stack to identify: (i) the identifiable subspace and (ii) its orthogonal blind complement. Then either:
\begin{itemize}
  \item Project the HNN update onto the identifiable subspace (drop updates along blind directions), or
  \item Selectively damp the blind directions with a small leakage toward a neutral prior (e.g., the last well-identified estimate), leaving the excited directions untouched.
  \end{itemize}
\end{itemize}

This monitor procedure preserves nominal behaviour when the problem is well excited, gently boosts curvature when conditioning degrades, and finally prevents drift by projecting or damping along blind directions, while leaving the dynamics unchanged where identifiability is adequate.

\section{Simulation Results}\label{sec:simulations}
\begin{figure}[!t]
	\centering
	\includegraphics[width=3.5in]{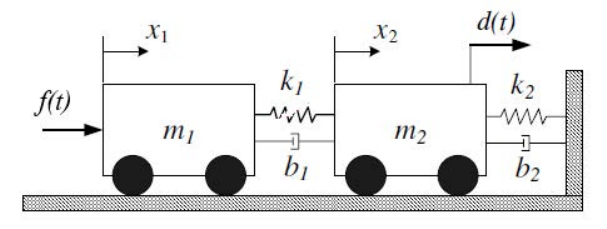}
	\caption{Model of the 2-DOF Mass–Spring–Damper (MSD) system.}
	\label{fig1}
\end{figure}

We evaluate the proposed HNN estimators on a 2-DOF mass–spring–damper (MSD) plant (Fig.~\ref{fig1}; cf.~\cite{FEKRI2006,FEKRI2007}) with unknown constants $k_1,b_1,k_2,b_2$, a disturbance $d(t) \sim \mathcal{N}(\mu,\sigma^2)$ (white Gaussian), and a slowly time–varying stiffness $k_1(t)$ and compare with the performance of PB-RLS, DA-PB-KF and DA-PB-MHE algorithms.
Subsection~B presents the behaviour of the standard LS (unconstrained) HNN. 
Subsections~C–E compare the proposed HNN estimators against PB–RLS (constraints only), DA–PB–KF (constraints+disturbances), and DA–PB–MHE (constraints+disturbances+time–varying parameters) on a common setup. 
Subsection~F reports a 10–trial Monte Carlo across varying initial conditions, disturbance realizations, and parameter drift. 
We conclude with a brief complexity comparison of all four methods.

\subsection{Dynamical System and Estimation Model}\label{subsec:MSE_model}
We use the 2-DOF mass–spring–damper (MSD) of Fig.~\ref{fig1} (cf.~\cite{FEKRI2006,FEKRI2007}): two masses
$m_1,m_2$ coupled by $(k_1,b_1)$, with mass~2 attached to ground by $(k_2,b_2)$. The first mass is actuated by a
known force $f$, and an additive disturbance $d$ may act on the second mass. Let $x=[x_1,x_2,\dot x_1,\dot x_2]^\top$
and $u=[f,d]^\top$. The masses $m_1,m_2$ are known; $(k_1,b_1,k_2,b_2)$ are unknown to be estimated. We consider
scenarios with $d(t)=0$, $d(t) \sim \mathcal{N}(\mu,\sigma^2)$ (constant $+$ white Gaussian noise) and a slowly time–varying $k_1(t)$.

\paragraph{Continuous-time plant}
\begin{subequations}\label{eq-52}
\begin{align}
\dot{x}&=A_cx+B_cu + H_c d, \label{eq-52a}\\
y&=C_cx+D_cu+E_c d, \label{eq-52b}
\end{align}
\end{subequations}
with
\begin{equation}\label{eq-53}
A_c=\begin{bmatrix}
0&0&1&0\\ 0&0&0&1\\ -\tfrac{k_1}{m_1}&\tfrac{k_1}{m_1}&-\tfrac{b_1}{m_1}&\tfrac{b_1}{m_1}\\
\tfrac{k_1}{m_2}&-\tfrac{k_1+k_2}{m_2}&\tfrac{b_1}{m_2}&-\tfrac{b_1+b_2}{m_2}
\end{bmatrix},\quad
B_c=\begin{bmatrix}
0\\ 0\\ \tfrac{1}{m_1}\\ 0
\end{bmatrix},\quad H_c=\begin{bmatrix}0 \\ 0 \\ 0 \\ \tfrac{1}{m_2}  \end{bmatrix}, \quad
C_c=I,\ D_c=E_c =0 .
\end{equation}

\paragraph{Linear-in-parameters regression}
From \eqref{eq-52a}, with measured (or estimated) accelerations but unmeasurable disturbance $d$ and without compensation neurons yields:
\begin{equation}\label{eq-54}
w=W\theta,
\end{equation}
where
\begin{subequations}\label{eq-55}
\begin{align}
\theta&=[\,\hat k_1,\hat b_1,\hat k_2,\hat b_2\,]^\top, \label{eq-55a}\\
w&=[\,m_1\ddot x_1-f,\ \ m_2\ddot x_2 - d\,]^\top, \label{eq-55b}\\
W&=\begin{bmatrix}
x_2-x_1 & -\dot x_1+\dot x_2 & 0 & 0\\
-x_2+x_1 & -\dot x_2+\dot x_1 & -x_2 & -\dot x_2
\end{bmatrix}. \label{eq-55c}
\end{align}
\end{subequations}

\paragraph{Discretisation and signals}
Both the MSD and the HNN estimator are integrated with fixed–step RK4 (step $h$) to avoid numerical artifacts and allow a practical $h$. The discrete plant reads
\begin{subequations}\label{eq-57}
\begin{align}
x(k{+}1)&=A_d x(k)+B_d u(k) + H_d d(k), \label{eq-57a}\\
y(k)&=C_d x(k)+D_d u(k) + E_d d(k), \qquad (C_d{=}I,\ D_d{=}E_d=0), \label{eq-57b}
\end{align}
\end{subequations}
and the discrete regression becomes
\begin{equation}\label{eq-58}
w(k)=W(k)\,\theta(k),
\end{equation}
with
\begin{subequations}\label{eq-59}
\begin{align}
\theta(k)&=[\,\hat k_1(k),\hat b_1(k),\hat k_2(k),\hat b_2(k)\,]^\top, \label{eq-59a}\\
w(k)&=\big[m_1 a_1(k)-f(k),\ m_2 a_2(k)-d(k)\big]^\top
\approx \Big[m_1\frac{v_1(k)-v_1(k\!-\!1)}{h}-f(k),\ m_2\frac{v_2(k)-v_2(k\!-\!1)}{h}-d(k)\Big]^\top, \label{eq-59b}\\
W(k)&=\begin{bmatrix}
x_2(k)-x_1(k) & -v_1(k)+v_2(k) & 0 & 0\\
-x_2(k)+x_1(k) & -v_2(k)+v_1(k) & -x_2(k) & -v_2(k)
\end{bmatrix}. \label{eq-59d}
\end{align}
\end{subequations}
(Accelerations via first–order backward differences).

\paragraph{Discrete HNN}
With $v=\alpha\tanh\!\big(\tfrac{\beta}{2}u\big)$ and the projector mapping of Section~\ref{sec:mapping},
\begin{subequations}\label{eq-60}
\begin{align}
u(k{+}1)&=u(k)+T_d(k)\,v(k)+b_d(k), \label{eq-60a}\\
v_i(k)&=\alpha\tanh\!\big(\tfrac{\beta}{2}u_i(k)\big),\quad i=1,\dots,n. \label{eq-60b}
\end{align}
\end{subequations}
We use the direct relation $v=\theta$ and $\alpha$ is chosen large enough to cover the parameter ranges (a neuron-range setting, not a hard constraint).

\paragraph{Simulation setup}
Unless stated otherwise: $m_1{=}m_2{=}1$; true parameters $k_1^*=1$, $b_1^*=0.15$, $k_2^*=0.5$, $b_2^*=0.25$ (thus $v^*=[k_1^*,b_1^*,k_2^*,b_2^*]^\top$); input
$f(t)=1+\sin(t)\cos(2t)+\cos(3t)+\sin(0.5t)$; initial state $x(0)=[0,\,0.3,\,0,\,0]^\top$; $a(0)=[0,0]^\top$; HNN initialisation $v(0)=[0.25,\,0.05,\,0.3,\,0.15]^\top$.

\subsection{Simulation of Standard LS–HNN (unconstrained)}
\begin{figure}[!t]
  \centering
  \includegraphics[width=5.5in]{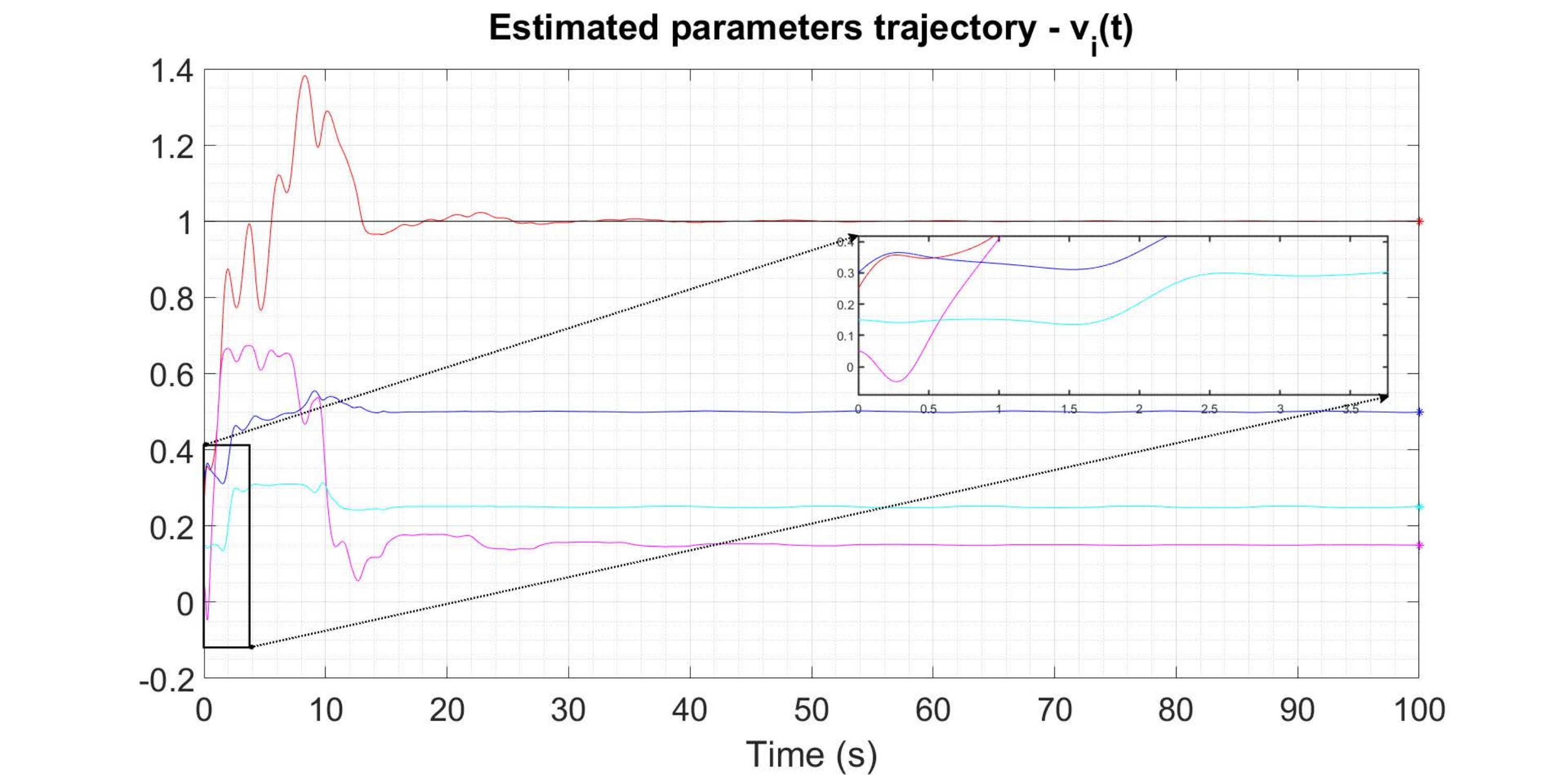}
  \caption{Standard LS–HNN (no constraints). Parameter trajectories:
  \(v_1{=}\hat k_1\) (red), \(v_2{=}\hat b_1\) (magenta), \(v_3{=}\hat k_2\) (blue), \(v_4{=}\hat b_2\) (cyan); Asterisks mark true values. LS-HNN settings: $h=10^{-4}$, \(\beta=1\), $\alpha=6$.}
  \label{fig2}
\end{figure}

\begin{figure}[!t]
  \centering
  \includegraphics[width=5.5in]{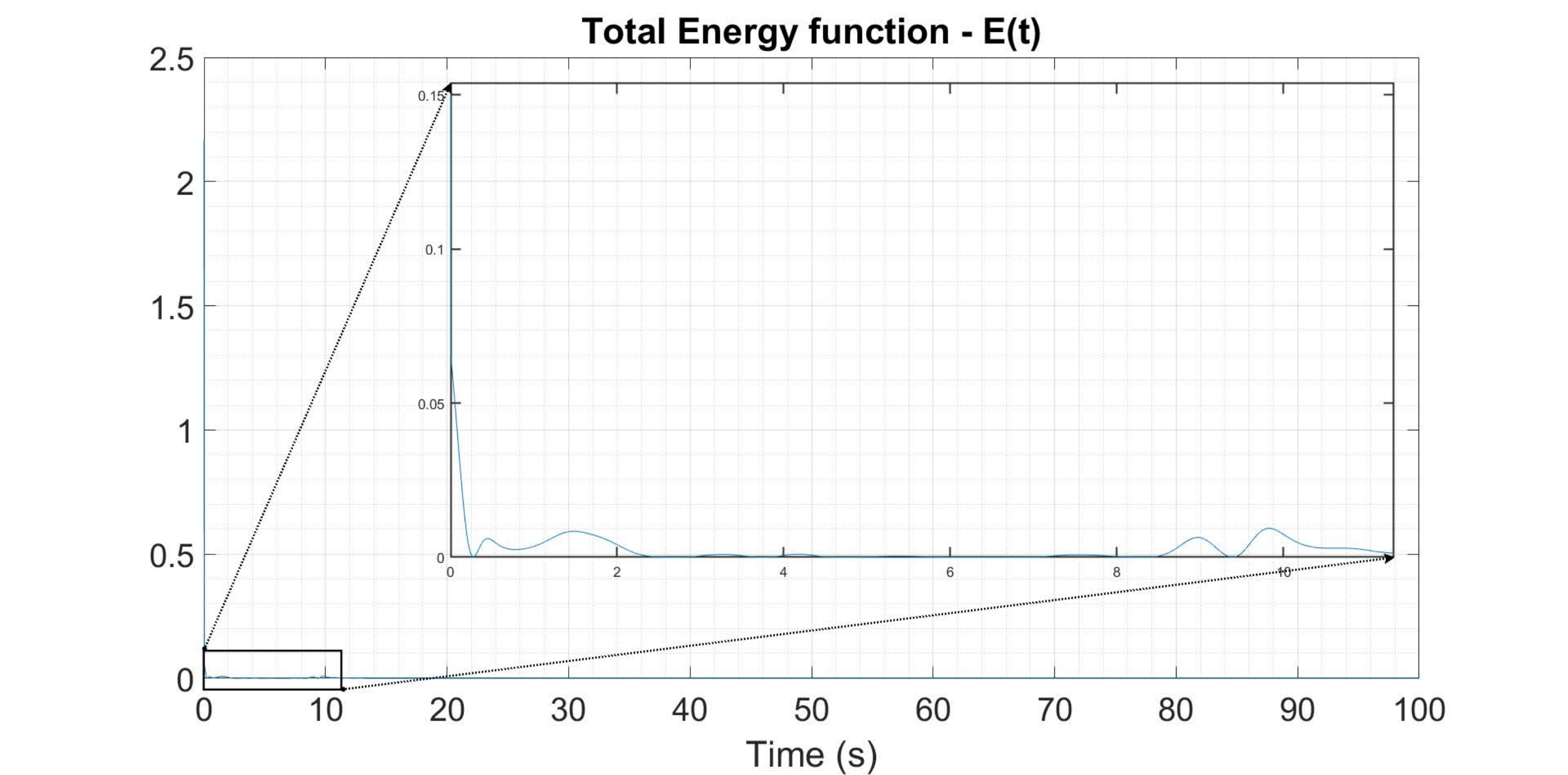}
  \caption{Standard LS–HNN: energy function \(E(t)\) for \(\beta=1\). Non-monotonic due to time-varying \(T,b\). HNN settings: $h=10^{-4}$, \(\beta=1\), $\alpha=6$}
  \label{fig3}
\end{figure}

We first implement the standard, unconstrained LS–HNN. 
Figure~\ref{fig2} shows a run with $h=10^{-4}$, $\alpha=6$ and \(\beta=1\). It was considered no disturbances (\(d=0\)), and constant true parameters. 
The estimates approach the correct values, but (i) \(\hat b_1\) becomes temporarily negative and is therefore physically infeasible, highlighting the need for constraints, and (ii) considerable oscillations appear in the transient. 
Figure~\ref{fig3} plots the energy function \(E(t)\), which is not monotone since \(T(t),b(t)\) vary with data, so \(E(t)\) is not a Lyapunov function under the standard mapping. 
These issues motivate the projector-based, constraint-aware HNN estimators presented next.

\subsection{Simulation of the Constrained Estimation Problem -- Proposed CA-HNN and PB-RLS}\label{subsec:sim_constr}

In subsequent analysis, the proposed CA-HNN estimator will be tested along with the PB-RLS algorithm. In the next simulations, the parameters will be subject to the following constraints: \(0.25 \leq k_1 \leq 1.75\), \(0.05 \leq b_1 \leq 0.25\), \(0.3 \leq k_2 \leq 0.7\) and \(0.15 \leq b_2 \leq 0.35\).

\begin{figure}[!t]
	\centering
    \hspace*{-1.5cm}
	\includegraphics[width=7.5in]{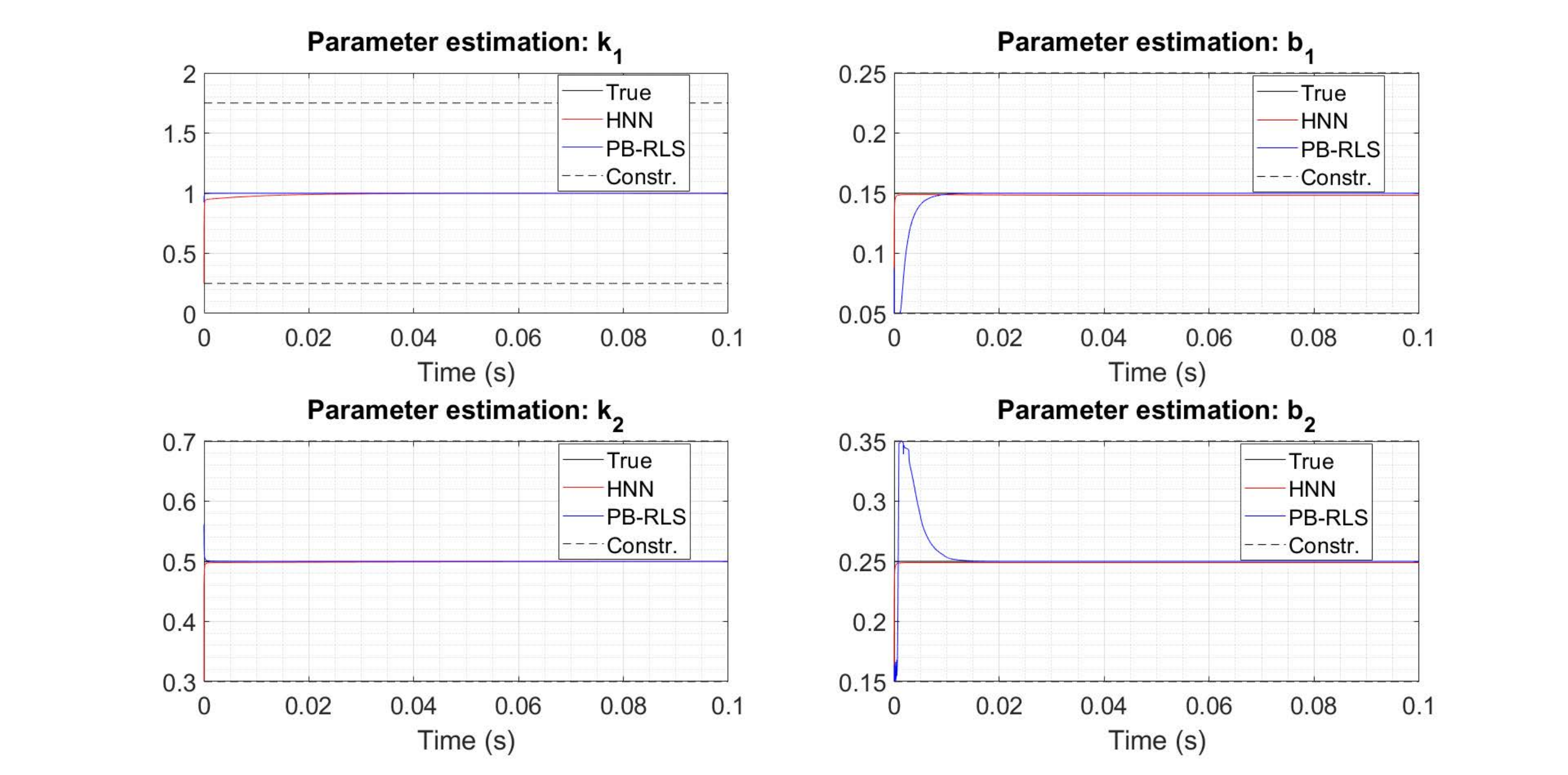}
    \caption{Comparison between proposed CA-HNN estimator and PB-RLS algorithm. MSD parameter estimates with constraints (constant parameters, no disturbance). CA-HNN settings: $h=10^{-5}$, $\alpha=10$, $\eta=50$ and $\beta=250$. PB–RLS settings: forgetting factor $\lambda=0.995$, initial covariance $P_0=10^6I$ and sampling period $h=10^{-5}$. CA-HNN converges smoothly and remains within bounds; PB--RLS reaches the same steady state but shows a transient overshoot for $b_2$ that briefly hits the upper bound.}
	\label{fig4}
\end{figure}

The comparison of the proposed CA-HNN estimator with the PB-RLS algorithm is presented in Fig. \(\ref{fig4}\). The CA-HNN estimator was simulated with parameters: $h=10^{-5}$, $\alpha=10$, $\eta=50$ and $\beta=250$. The PB–RLS algorithm (constraints-only) used: forgetting factor $\lambda=0.995$, initial covariance $P_0=10^6I$ and sampling period $h=10^{-5}$. The numerical results of ten Monte Carlo runs, covering different initial conditions, are presented in Subsection~\ref{numcomp}.

As we can see, both estimators converge to the true parameters within the window.
The proposed CA-HNN shows smooth, monotone transients for all parameters and stays strictly inside
the bounds, while PB--RLS achieves the same steady-state accuracy but exhibits a transient
overshoot for $b_2$ that touches the upper bound. These traces highlight the benefit of the constraint-aware projector mapping of the CA-HNN: fast, well-damped
convergence without boundary interaction.

\subsection{Simulation of the Constrained Estimation Problem Under Disturbances -- Proposed CA$^2$-HNN and DA-PB-KF}\label{subsec:sim_constr_distur}

Next, we present a simulation run when the MSD system is subject to constraints but also subject to a disturbance force \(d\). The disturbance force is assumed to be white Gaussian noise, with mean=1 and variance=1, i.e., $d(t) \sim \mathcal{N}(1,\,1)$.

\begin{figure}[!t]
	\centering
    \hspace*{-1.5cm}
	\includegraphics[width=7.5in]{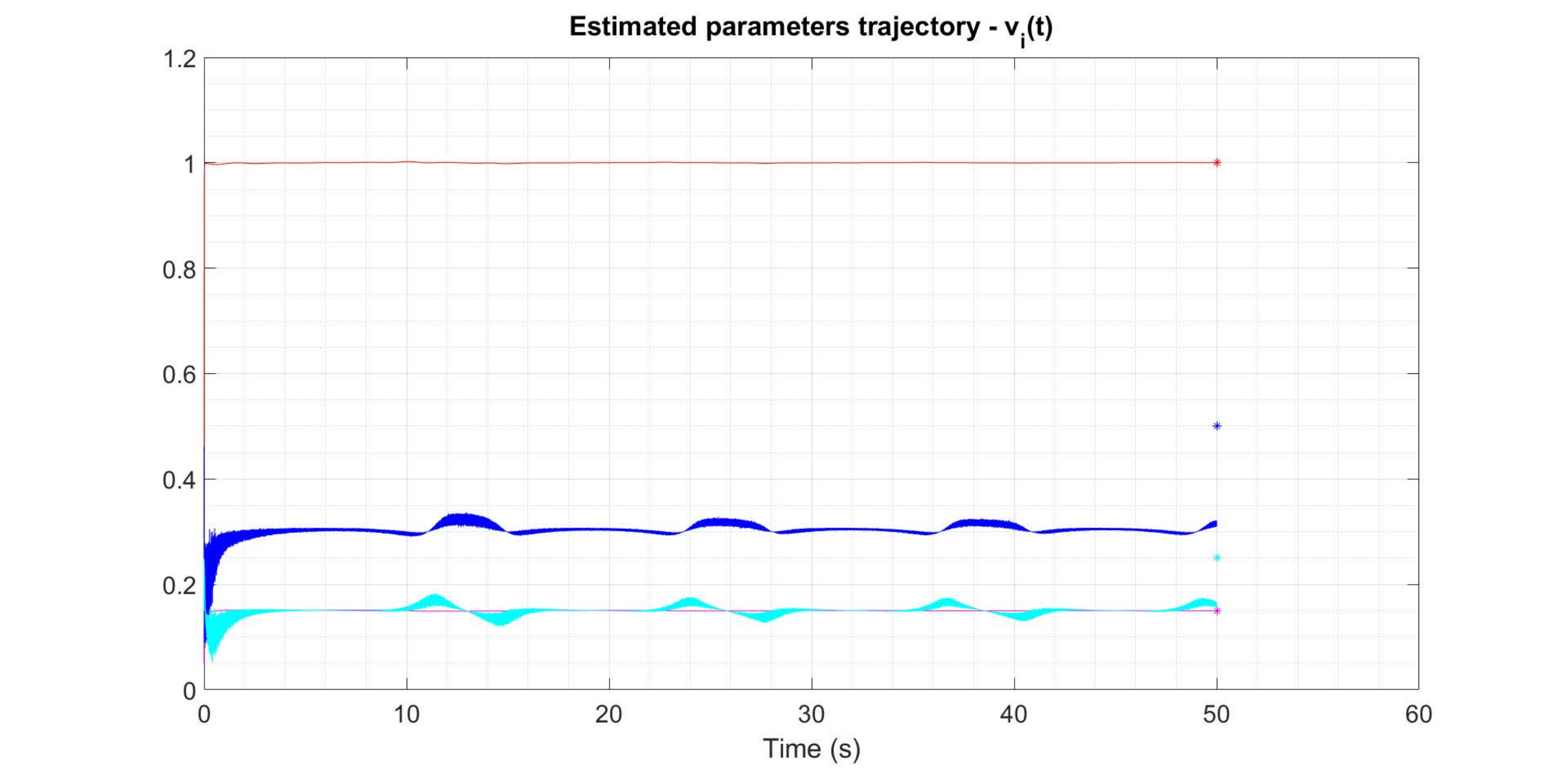}
	\caption{Simulation of CA-HNN estimator. Estimated parameters $v_i(t)$ under a disturbance $d(t) \sim \mathcal{N}(1,\,1)$ on $m_2$
(without compensation neurons). Asterisks mark true values.
$\hat k_1$ and $\hat b_1$ converge near their true values; $\hat k_2$ and $\hat b_2$
show a steady bias and a small ripple at the $d\!\to\!x_2$ resonance
(period $\approx 13.7$\,s). CA-HNN settings: $h=10^{-5}$, $\alpha=10$, $\eta=50$ and $\beta=200$. All constraints are satisfied at all times.}
	\label{fig5}
\end{figure}

\begin{figure}[!t]
	\centering
	\includegraphics[width=7.0in]{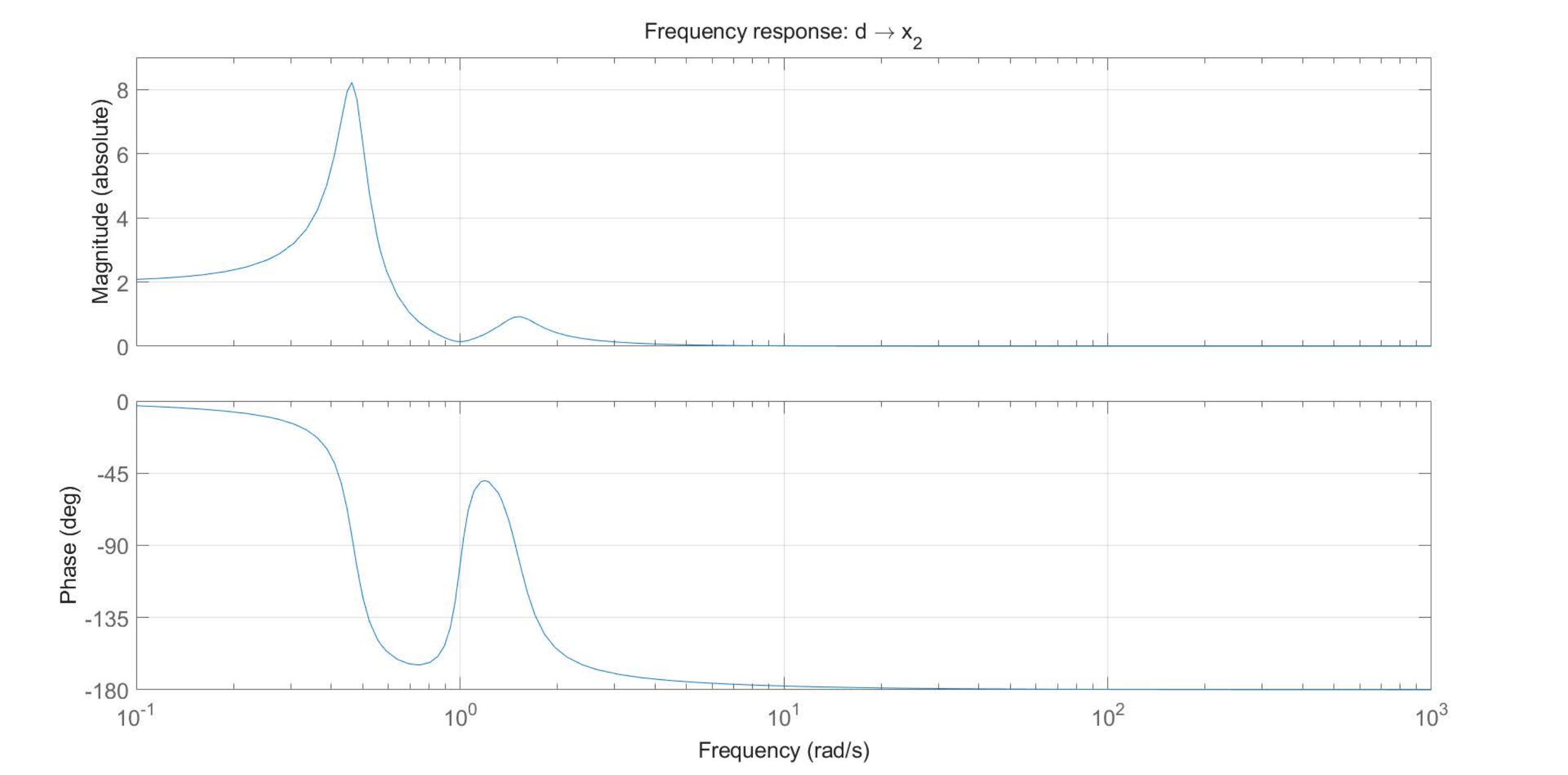}
    \caption{Bode magnitude (absolute units) and phase (degrees) of the transfer function \(G(j\omega)=X_2(j\omega)/D(j\omega)\) for the MSD system.}
	\label{fig6}
\end{figure}

First, we test the proposed CA-HNN estimator (without compensation neurons $v_d$) (see Fig. \ref{fig5}),
with the disturbance acting on $m_2$ and without compensation neurons.
Estimates $\hat k_1=v_1$ (red) and $\hat b_1=v_2$ (magenta) converge quickly and remain
close to their true values ($\approx 1.0$ and $0.15$), showing small, well–damped transients.
In contrast, the parameters tied to the second mass, $\hat k_2=v_3$ (blue) and
$\hat b_2=v_4$ (cyan), exhibit (i) a clear steady–state bias and (ii) a small
quasi–periodic ripple. At $t{=}50$\,s the bias is notable
($\hat k_2\!\approx\!0.30$ vs.\ $0.50$, $\hat b_2\!\approx\!0.14$ vs.\ $0.25$),
because the additive term $d$ enters the $d\!\to\!x_2$ channel and is absorbed by these
parameters in the absence of disturbance compensation. The ripple frequency matches the
plant resonance (cf.\ the $d\!\to\!x_2$ Bode diagram in Fig.~\ref{fig6}, peak near
$\omega_r\!\approx\!0.46$\,rad/s, period $\approx 13.7$\,s), confirming that the oscillation
is plant–induced rather than an HNN artifact. In fact, comparing
Figs.~\ref{fig5} and~\ref{fig6}, we note that the oscillation in the CA-HNN parameter estimates occurs at this same frequency. As shown next, adding compensation
neurons removes the bias by channelling
unmodeled additive effects into $v_d$, and a reduction of
$\beta$ attenuates the residual ripple. 

Hence, to mitigate the effect of additive disturbance, which is not measurable, it is proposed to add another neuron \(v_d\) to the estimation model, which will compensate for the effect of the disturbance on the regression model. This neuron is constrained to belong to the interval \( ] - \alpha , + \alpha [ \). Therefore, as presented in Subsection~\ref{compensation}, the estimation model presented in (\ref{eq-54})-(\ref{eq-55}) must be modified to:
\begin{equation}\label{eq-61}
	w =W_{aug} v_{aug}			 
\end{equation}
where,
\begin{subequations}\label{eq-62}
	\begin{align}
		v_{aug}&=\left[ \hat{k}_1, \hat{b}_1, \hat{k}_2, \hat{b}_2 , v_d \right ]^T   \label{eq-62a}\\
		w&= \left[ m_1\ddot{x}_1-f, m_2\ddot{x}_2 - d \right]^T  \label{eq-62b}\\
		W_{aug}&= \left[ \begin{array}{ccccc}
			x_2 -x_1 & -\dot{x}_1+ \dot{x}_2 & 0 & 0 & 0 \\
			-x_2+x_1 & -\dot{x}_2+\dot{x}_1 & -x_2 & -\dot{x}_2 & 1
		\end{array} \right] \label{eq-62c}
	\end{align}
\end{subequations}

\begin{figure}[!t]
	\centering
	\includegraphics[width=4.5in]{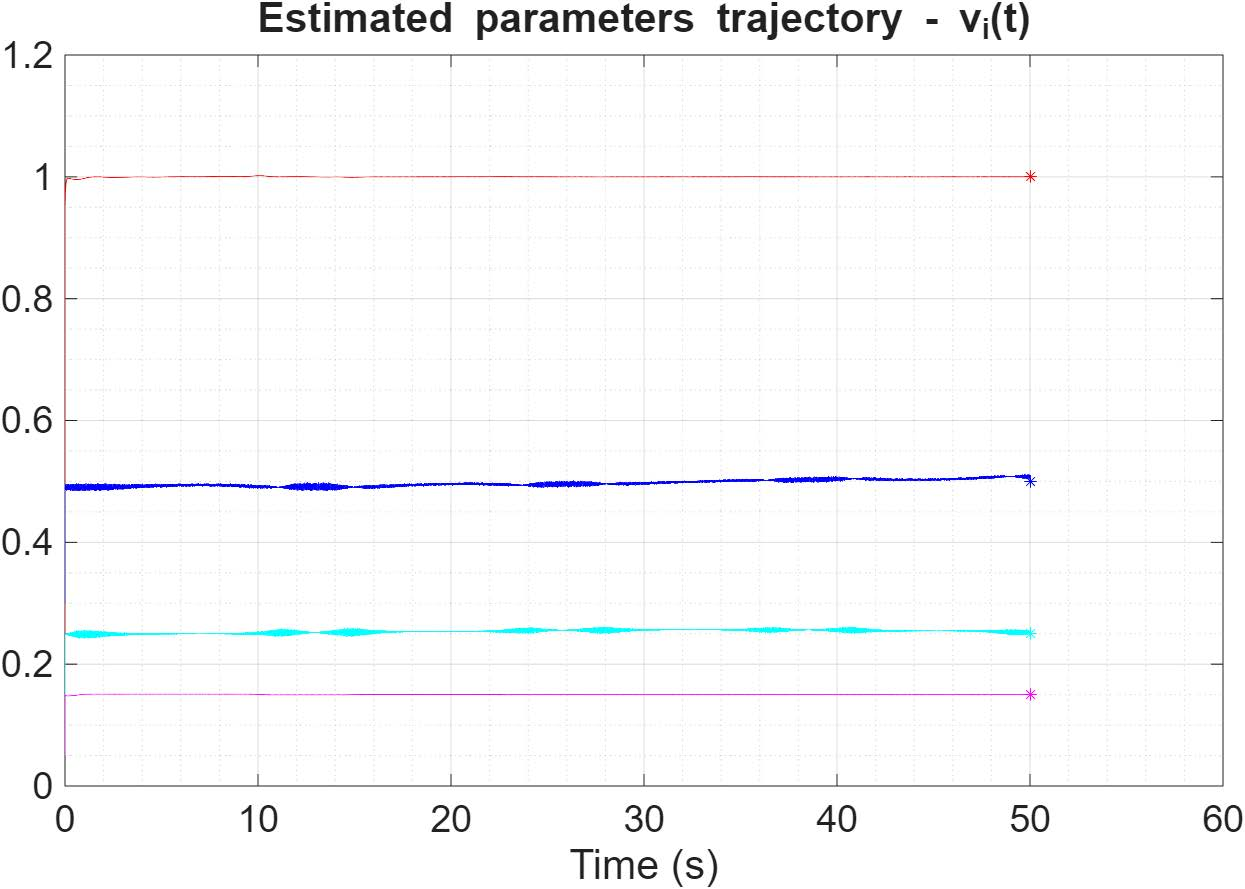}
	\caption{Trajectories of the estimated parameters with the compensation neuron $v_d$ enabled.
Colours: $\hat{k}_1=v_1$ (red), $\hat{b}_1=v_2$ (magenta), $\hat{k}_2=v_3$ (blue), and $\hat{b}_2=v_4$ (cyan).
Disturbance: white Gaussian noise with mean $1$ and variance $1$.
CA$^2$-HNN settings: step $h=10^{-5}$, saturation $\alpha=10$, constraint weight $\eta=50$, and gain $\beta=100$.
Estimates settle near their targets with small ripple; Asterisks mark reference values. All constraints are satisfied at all times.}
	\label{fig7}
\end{figure}

The simulation result, with compensation neuron enabled, is presented in Fig. (\ref{fig7}). In fact, with the compensation neuron $v_d$ enabled and an additive disturbance $d(t) \sim \mathcal{N}(1,\,1)$, the four parameter estimates remain essentially constant and close to their reference values (asterisks), showing no steady bias. The small ripples visible on each trace reflect the white Gaussian component of the disturbance and the chosen moderate bandwidth. Constraints keep the trajectories feasible without chatter. This behavior matches the analysis for the augmented estimator: the disturbance is absorbed by $v_d$, while the parameter block contracts under $P_{\mathrm{WH}}+\eta P_{A,\mathrm{aug},\theta}$, yielding an unbiased steady regime. With the selected settings ($h=10^{-5}$, $\alpha=10$, $\eta=50$, $\beta=100$), the tracking bandwidth scales as $\gamma\!=\!(\alpha\beta/2)\,\delta\,c_{\mathrm{aug}}$ and the ultimate error as $\rho\!\propto\!(\alpha\beta\,c_{\mathrm{aug}}^2)^{-1/2}$, explaining the smooth (low-noise) but slightly conservative dynamics. Increasing $\beta$ would speed convergence at the cost of higher high-frequency ripple, while larger $\alpha$ reduces saturation risk but must respect the discrete-time stability goal.

\begin{figure}[!t]
	\centering
    \hspace*{-1.5cm}
	\includegraphics[width=7.5in]{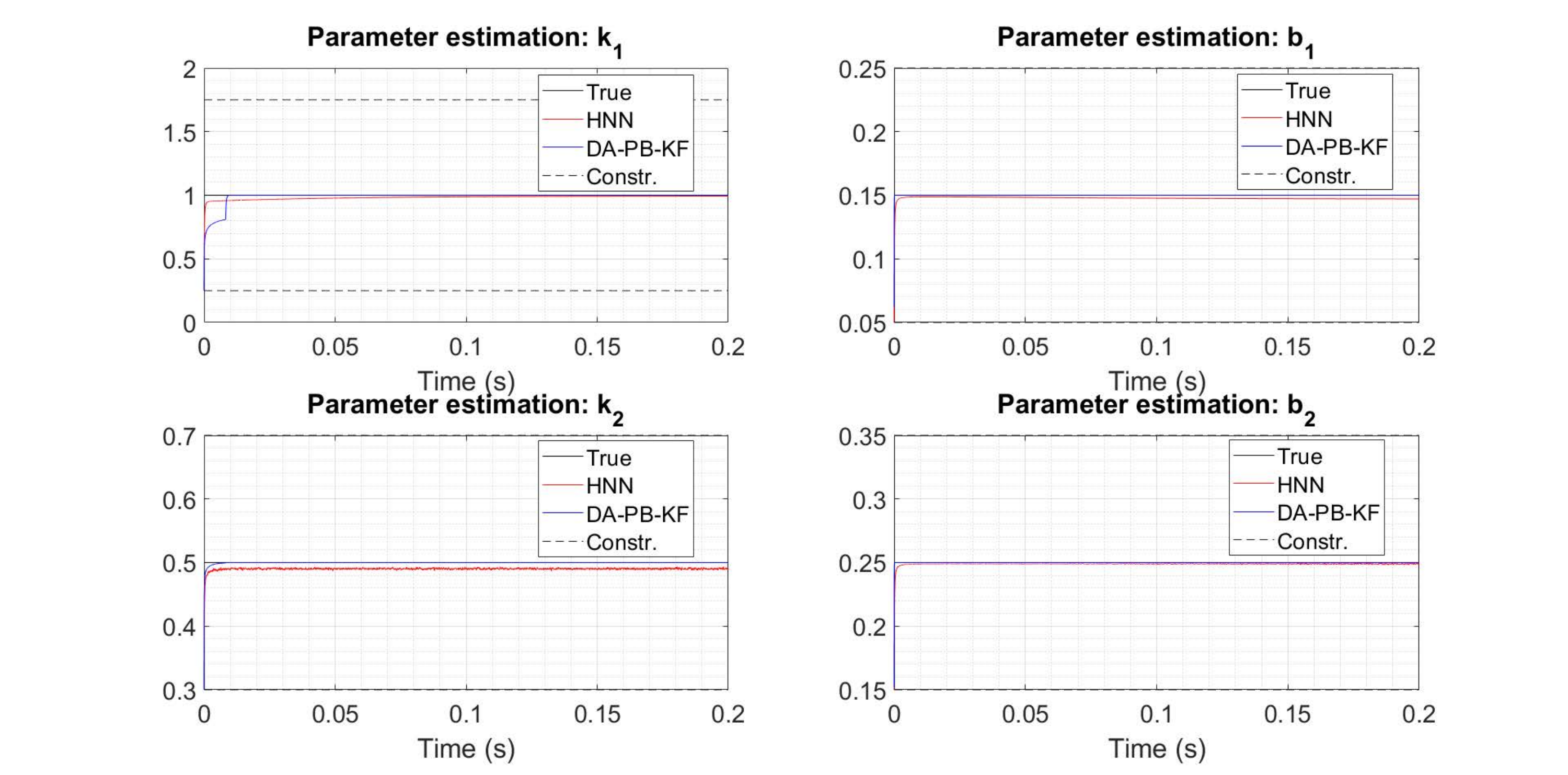}
    \caption{Comparison of parameter estimates for the MSD model with additive disturbance ($d(t) \sim \mathcal{N}(1,\,1)$):
CA$^2$-HNN (red) vs.\ DA–PB–KF (blue), with true values (black) and box constraints (dashed). CA$^2$-HNN settings: step $h=10^{-5}$, saturation $\alpha=10$, constraint weight $\eta=50$, and gain $\beta=50$. DA–PB–KF settings: 
$Q_d=10^{12}$, $R=10^{12} I_2$, $P_{0,\theta}=10^{6} I_4$, projection onto box constraints after each update. All four parameters (\(k_1,b_1,k_2,b_2\)) converge rapidly (\(<\!0.02\,\mathrm{s}\)) to the
true values and remain within the feasible set. The DA–PB–KF shows a small initial
step on \(k_1\) due to the diffuse prior and constraint projection, while the CA$^2$-HNN
transient is monotone; steady-state accuracy is essentially identical for both.}
	\label{fig8}
\end{figure}

Figure~\ref{fig8} compares the estimation of the four MSD
parameters $(k_1,b_1,k_2,b_2)$ under additive disturbance, $d(t) \sim \mathcal{N}(1,\,1)$, for two estimators:
the proposed CA$^2$-HNN (red) and DA–PB–KF (blue). The black lines mark the true
values and the dashed lines indicate the box constraints used in both methods.
Both estimators receive the same measurements/regressor and enforce the same
constraints.
All parameters converge rapidly (within a few tens of milliseconds) to the true
values and stay inside the feasible set, showing unbiased steady performance for
both algorithms. The only visible difference is an initial step on $k_1$ for
DA–PB–KF: this is expected with a diffuse prior and constraint projection, which
produce a large first Kalman gain and an immediate pull to the feasible region.
The CA$^2$-HNN transient is monotone and of comparable speed; minor ripples are disturbance–induced. Overall, the two approaches achieve essentially
the same accuracy and settling time under the same constraints and disturbance
modelling. The numerical results of ten independent Monte Carlo runs, with varying disturbance means and variances, are presented in Subsection~\ref{numcomp}.

\subsection{Simulation of the Constrained Estimation Problem with Time–Varying Parameters under Disturbances—Proposed CA$^2$-HNN vs.\ DA–PB–MHE}\label{subsec:sim_constr_distur_param}

We compare the proposed CA$^2$-HNN against a disturbance-augmented, projection-based moving-horizon estimator (DA--PB--MHE) on the MSD system subject to (i) box constraints on the physical parameters, (ii) an additive disturbance, and (iii) a slowly time-varying parameter \(k_1(t)\).

\begin{figure}[!t]
	\centering
    \hspace*{-1.5cm}
	\includegraphics[width=7.5in]{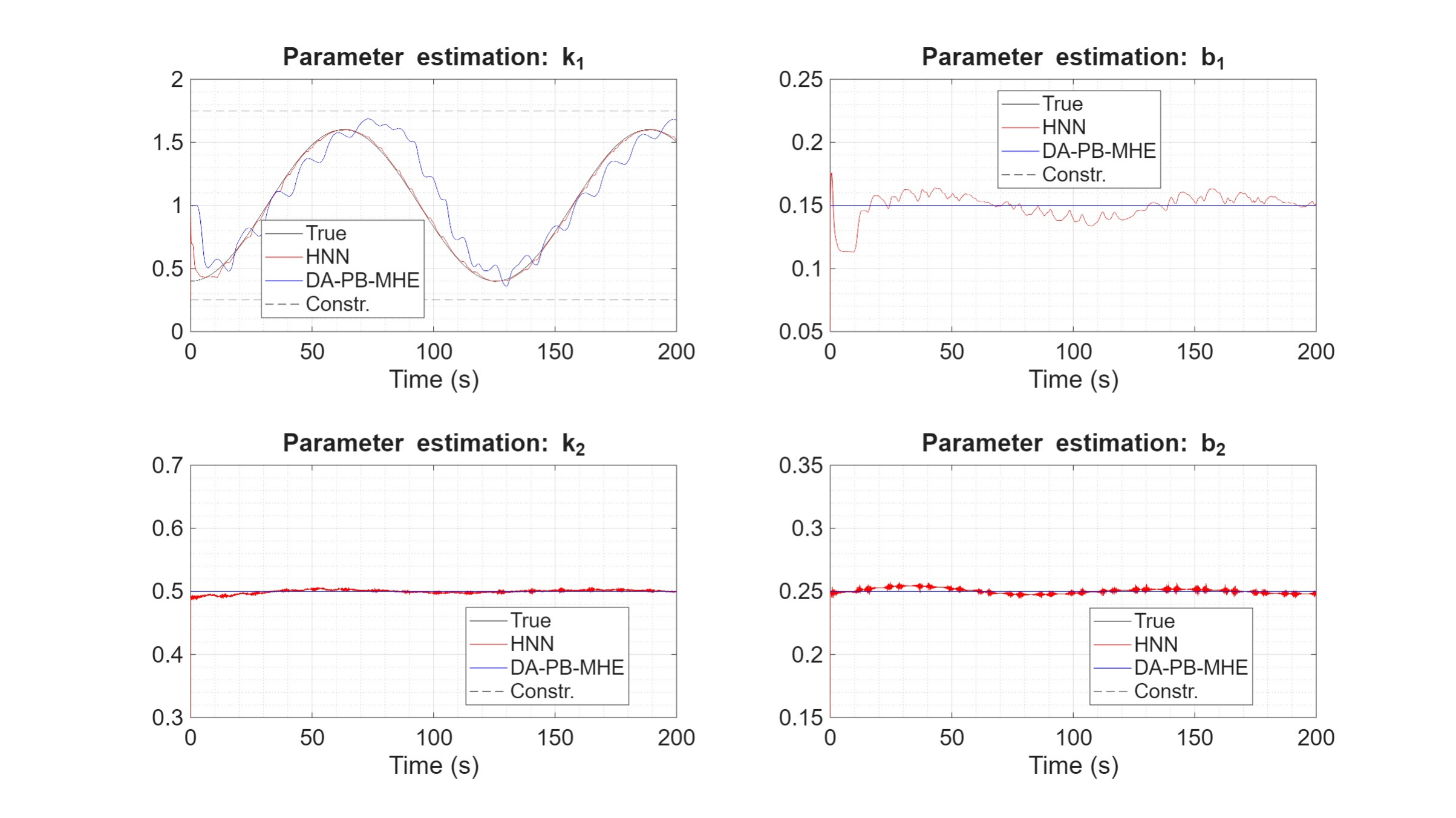}
\caption{Comparison of parameter estimates for the MSD model with constraints, additive disturbance ($d(t) \sim \mathcal{N}(1,\,1)$) and time-varying parameter ($k_1(t)=1-0.6\cos(0.05\times t)$):
CA$^2$-HNN (red) vs.\ DA–PB–MHE (blue), with true values (black) and box constraints (dashed). CA$^2$-HNN settings: step $h=10^{-4}$, saturation $\alpha=10$, constraint weight $\eta=50$, and gain $\beta=10$. DA–PB–MHE main settings: $h=10^{-4}$, decimation \(M=100\), sliding horizon \(L_{\mathrm{win}}=20\), velocity IIR prefilter \(\alpha=0.9\), box constraints (dashed), \(Q_\theta=\mathrm{diag}(9\!\times\!10^{-4},0,0,0)\), \(P_{0,\theta}=\mathrm{diag}(9,10^{-8},10^{-8},10^{-8})\), \(R=(10^{3})^{2}I_{2}\), AR(1) with \(\rho=0.98\), innovation st.\ \(\sigma_w=5\), and arrival \ \(\sigma_{d0}=50\) (no explicit bounds on \(d\)) and 
constrained least-squares backend (quadprog).
 For \(k_1\), CA$^2$-HNN tracks the time-varying parameter with low lag and little ripple, while DA–PB–MHE shows mild oscillation/overshoot around fast transients. 
For \(b_1\), both methods converge near the true value with small steady ripples. 
For \(k_2\) and \(b_2\), both estimators remain essentially constant at the correct values and respect the bounds throughout, with CA$^2$-HNN slightly smoother overall.}
	\label{fig9}
\end{figure}

For the DA--PB--MHE implementation, we use a sliding window of length \(N\) on decimated data (decimation \(M\); estimator step \(\Delta t=Mh\)), with an arrival cost at the window head. Constraints are enforced via a projection-based map, and the model is augmented with a scalar disturbance state \(d_t\) that follows a first-order autoregressive (AR-(1)) dynamics \(d_t=\rho d_{t-1}+w_t\) to capture bias/coloured effects. The MHE cost comprises a measurement term with covariance \(R\), parameter-drift regularisation with process covariance \(Q_\theta\) (allowing \(k_1\) to vary and keeping the other parameters nearly constant within the window), and disturbance regularisation with \(Q_d\); an arrival covariance \(P_0\) initialises the window. Numerically, we apply a small diagonal ridge to ensure positive definiteness, warm-start the solver from the previous solution, and solve each step with a constrained least-squares/QP backend (Gauss--Newton/LM with trust-region damping). A light IIR prefilter on velocities stabilises the residuals. (Concrete values for \(N,M,R,Q_\theta,Q_d,P_0,\rho\) etc., are given in the Fig.~\ref{fig9} caption.)

The proposed CA$^2$-HNN requires only a handful of gains and scales (\(\alpha,\beta,\eta\), step size \(h\)), plus the constraint matrix. Disturbance handling is embedded via compensation neurons without new tuning. In contrast, DA--PB--MHE exposes many coupled parameters---horizon \(N\), decimation \(M\), \(R\), \(Q_\theta\), \(Q_d\), arrival \(P_0\), first-order autoregressive disturbance parameters \(\rho,\sigma_w,\sigma_{d0}\), solver damping and tolerances, and numerical ridge. This larger design space can yield excellent performance when carefully tuned, but it increases tuning effort and computational load. Practically, DA--PB--MHE incurs a per-step optimisation (time-consuming at high rates), whereas the CA$^2$-HNN update is a small set of matrix--vector operations; both are usable online, but MHE typically demands more compute or decimation to meet real-time budgets.
Fig.~\ref{fig9} shows that for all parameters, both estimators respect the imposed bounds (dashed lines) for the entire run, i.e., feasibility is invariant. For \(b_1\) and \(k_2, b_2\), both methods remain close to the true values. For the time–varying \(k_1(t)\), the proposed CA$^2$-HNN tracks the true values with low lag and low ripple, whereas the DA–PB–MHE trace exhibits small oscillations and some lag. The ripple in \(k_1(t)\), estimated by DA-PB-MHE, occurs because the AR(1) disturbance model is a low-pass filter. When the plant is excited near its resonance, the residual has a pronounced sinusoidal component. Because (i) the AR(1) disturbance model penalises that oscillation and (ii) since the sensitivity of \(x_2\) to \(k_1\) is maximal near resonance, the optimiser preferentially decreases the residual by allowing a slight oscillation of \(k_1\) at the resonant frequency. The sliding finite horizon amplifies this trend: as the window moves, the same narrowband energy re-enters and exits the fit, causing a repeating rebalancing and a visible ripple in \(k_1(t)\).

In the proposed CA$^2$-HNN, a compensation neuron is introduced as an additive disturbance state coupled through the augmented regressor \([W\; H]\).
For a fixed parameter vector \(\theta\), the optimal value of this state coincides with the instantaneous residual \(w - W\theta\), i.e.,
\(d^*(t)=w(t)-W(t)\theta(t)\).
Because no temporal prior or feasibility constraints are imposed on the compensation neuron, it can track residual components across the full bandwidth of interest.
Inequality constraints are enforced solely on the physical parameters, not on the compensation neuron, so the constraint projector does not penalise this channel.
Consequently, narrowband residual energy near the resonance of the plant is absorbed by the compensation pathway rather than accommodated by oscillatory variations of \(k_1\).
Since this pathway is decoupled from parameter regularisation and constraint curvature, the CA$^2$-HNN does not have an incentive to encode the sinusoidal component in \(k_1\), thus suppressing ripple while preserving parameter feasibility. 

One possible mitigation for the ripple in DA--PB--MHE is to endow the disturbance with a harmonic AR(2) prior tuned near the resonant frequency, which can absorb narrowband energy and reduce leakage into \(k_1\).
However, to ensure a fair baseline and avoid ad-hoc tailoring to this specific system, we deliberately retain the canonical DA--PB--MHE configuration with an AR(1) disturbance prior in all comparisons.

A Monte Carlo study over ten different frequencies of $k_1(t)$ is reported in Subsection~\ref{numcomp}.

\subsection{Comparative Numerical Experiments}\label{numcomp}

Next, we evaluate, based on 10 Monte Carlo trials, the proposed HNN estimators against PB–RLS (constraints only), DA–PB–KF (constraints+disturbances), and DA–PB–MHE (constraints+disturbances+time–varying parameters). The case study is the 2-DOF MSD model, previously defined, with
linear-in-parameters regression, box constraints on physical parameters, additive unknown disturbances,
and slow time-varying stiffness $k_1(t)$.

Each estimator (CA-HNN / CA$^2$-HNN, PB–RLS, DA–PB–KF, DA–PB–MHE) is run on the same
synthetic data per trial to enable fair, paired comparisons. Algorithmic
hyper–parameters are fixed a priori and kept constant across all runs.
For each scenario we reuse exactly the hyper–parameters reported in the
single–run studies: Scenario~S1 (constraints only) uses the configurations
from Subsection~\ref{subsec:sim_constr};  Scenario~S2 (constraints\,+\,disturbance) uses those
from Subsection~\ref{subsec:sim_constr_distur} ; and Scenario~S3 (constraints\,+\,disturbance\,+\,time–varying
parameter) uses those from Subsection~\ref{subsec:sim_constr_distur_param}. This includes the same pre–filter
and decimation, horizons, gains, weights, and constraint penalties specified
therein. No retuning is performed across Monte Carlo trials; only the
randomised elements described next change.

We evaluate three scenarios. In all cases the constraints are those given in Subsection~\ref{subsec:sim_constr} and the true parameter values for $(k_1, b_1, k_2, b_2)$ are given in Subsection~\ref{subsec:MSE_model}.
\begin{itemize}
  \item[S1.] \emph{Random initial conditions}. Estimator states are drawn
  independently at each trial from the admissible box:
  $\hat\theta_i(0)\sim\mathcal{U}[\,\ell_i,u_i\,]$;
  \item[S2.] \emph{Additive disturbance}. The plant is driven by a white
  Gaussian disturbance; we use
  $d(t)\sim\mathcal{N}(\mu_d,\sigma_d^2)$ with the realization re–drawn
  at each Monte Carlo trial. Initial conditions are fixed to $\theta(0)=[0.25,\,0.05,\,0.3,\,0.15]^\top$. $\mu_d \in [1,5]$ and $\sigma^2_d \in [1,10]$.
  \item[S3.] \emph{Time–varying parameter.} The stiffness $k_1(t)$ follows
  a smooth trajectory of the form $k_1(t)=1-0.6\cos(\omega t)$; the frequency $\omega$ is drawn uniformly from a prescribed range
  $[\omega_{\min},\omega_{\max}]=[0.01,1]$ at each trial, while the disturbance and initial conditions are fixed to $d(t)\sim\mathcal{N}(1,1)$ and $\theta(0)=[0.25,\,0.05,\,0.3,\,0.15]^\top$.
\end{itemize}

For each scenario and parameter estimator, we report in Table~\ref{tab:agg} the following metrics:
\begin{enumerate}
  \item \emph{Final MSE}: the Mean–Square Error (MSE) over the last $10\%$ of the run,
  \[
  \mathrm{MSE}^{\mathrm{final}}_i
  \;=\;\frac{1}{K_{\!f}}\sum_{k=K-K_f+1}^{K} e_i[k]^2, \quad K_f=\lfloor 0.1K\rfloor .
  \]
  \item \emph{Area under the MSE–vs–time curve (AUC–MSE)}:
  discrete integral of the MSE over the whole run,
  \[
  \mathrm{AUC\text{-}MSE}_i
  \;=\;\Delta t\sum_{k=1}^{K} e_i[k]^2 .
  \]
  \item \emph{Settling times to $5\%$ and $1\%$ of error}: Define the normalised error, $e^{\mathrm{n}}_i[k] \;=\;
\frac{|e_i[k]|}{\theta^{\max}_i-\theta^{\min}_i}$.
  The settling time to $\varepsilon$ is defined as: $\text{Time}\!\to\!\varepsilon:
  \ \ \min\{\,t_k:\ e^{\mathrm{n}}_i[j]\le \varepsilon\ \ \forall j\ge k\,\},
  \quad \varepsilon\in\{0.05,0.01\}$, with a $1$\,s dwell (the condition must hold for at least $1$\,s) to
  avoid counting transient recrossings.
  \item \emph{Maximum constraint violation (\%)}:
  for box constraints $\ell_i\le\hat\theta_i\le u_i$,
  \[
  \mathrm{Viol}_{\max}(\%)
  = 100\!\times\!
  \max_k \max_i
  \frac{\big[\hat\theta_i[k]-u_i\big]_+ + \big[\ell_i-\hat\theta_i[k]\big]_+}
       {u_i-\ell_i}.
  \]
\end{enumerate}

All entries in Table~\ref{tab:agg} are reported as mean$\pm$std over
$N=10$ Monte Carlo trials. In S1 the initial conditions are randomised;
in S2 the disturbance realisations are randomised; in S3 the parameter
frequency $\omega$ is randomised. The same random seed is used for all
methods within each trial to ensure paired comparisons.

For time-varying parameters, all metrics are computed against the instantaneous truth. For parameter $i$
(with true trajectory $\theta_i[k]$ and estimate $\hat\theta_i[k]$ at the
estimator grid),
\[
e_i[k]=\hat\theta_i[k]-\theta_i[k],\qquad
\mathrm{MSE}_i=\frac{1}{K}\sum_{k=1}^{K} e_i[k]^2,\qquad
\mathrm{AUC\text{-}MSE}_i=\Delta t\sum_{k=1}^{K} e_i[k]^2 .
\]
Normalisation uses a scale that is robust for constant and varying parameters:
\[
c_i \;=\; \max\!\big\{\theta_i^{\max}-\theta_i^{\min},\; u_i-\ell_i\big\},\qquad
e_i^{\mathrm{n}}[k] \;=\; \frac{|e_i[k]|}{c_i},
\]
where $\theta_i^{\max/\min}$ are taken over the run and $[\ell_i,u_i]$ is the
box for $\theta_i$. (If $\theta_i$ is constant, $c_i$ defaults to the box
width.) Final MSE is computed over the last fraction of the run,
$\mathrm{MSE}^{\mathrm{final}}_i=\frac{1}{K_f}\sum_{k=K-K_f+1}^{K} e_i[k]^2$,
with $K_f=\lfloor 0.1K\rfloor$. For time–varying parameters this reflects the
steady tracking error (including any phase lag). Settling times to a band are defined as tracking times to an
\(\varepsilon\)-tube around the moving target:
\[
\mathrm{Time}\!\to\!\varepsilon
=\min\{\,t_k:\ e_i^{\mathrm{n}}[j]\le\varepsilon\ \ \forall j\ge k\,\},
\quad \varepsilon\in\{0.05,0.01\},
\]
with a 1\,s dwell requirement to avoid recrossings. This measures how quickly
the estimator locks onto the time–varying trajectory within a prescribed error
band. Constraint violation (\%) is calculated exactly as in the constant case
(supremum of normalised infeasibility over time), since constraints do not
depend on whether the parameters vary.

\begin{table}[t]
\centering
\scriptsize
\setlength{\tabcolsep}{4pt}
\caption{Monte Carlo comparison on the mass-spring-damper study ($N{=}10$). Mean$\pm$Std across trials.}
\begin{tabular}{llccccc}
\toprule
Scenario & Method & Final MSE & AUC--MSE & Time$\to$5\% [s] & Time$\to$1\% [s] & Constr. viol.\,[\%] \\
\midrule
\multirow{2}{*}{S1: Constraints only}
  & CA-HNN      & 1.3e-4$\pm$1.1e-4 & 1.5e-4$\pm$1.1e-4 & 5.1e-3$\pm$9.1e-5 & 1.7e-2$\pm$7.6e-3 & 0.0$\pm$0.0 \\
  & PB--RLS  & 7.2e-4$\pm$3.2e-6 & 7.2e-4$\pm$1.6e-6 & 1.1e-2$\pm$0.0 & 1.9e-2$\pm$0.0 & 0.0$\pm$0.0 \\
\addlinespace
\multirow{2}{*}{S2: + Disturbance}
  & CA$^2$-HNN        & 8.8e-4$\pm$5.7e-4 & 9.0e-4$\pm$5.6e-4 & 1.4e-2$\pm$1.2e-2 & ---- & 0.0$\pm$0.0 \\
  & DA--PB--KF & 2.4e-3$\pm$4.5e-6 & 2.4e-3$\pm$4.1e-6 & 1.4e-2$\pm$2.1e-5 & 1.4e-2$\pm$2.1e-5 & 0.0$\pm$0.0  \\
\addlinespace
\multirow{2}{*}{S3: + Time-varying params}
  & CA$^2$-HNN      & 1.2e-1$\pm$1.2e-1 & 1.4e-1$\pm$3.5e-2 & 9.5e+0$\pm$1.2e+1 & ---- & 0.0$\pm$0.0\\
  & PB--MHE  & 2.6e-1$\pm$2.6e-1 & 1.6e-1$\pm$7.8e-2 & ----- & ---- & 0.0$\pm$0.0 \\
\bottomrule
\end{tabular}
\label{tab:agg}
\vspace{5pt}
\begin{minipage}{0.97\linewidth}
\footnotesize
\textit{Legend (columns 3–7):}\\
(3) Final MSE — final mean squared deviation of parameter estimates (averaged over all parameters) at the end of the run. \\
(4) AUC--MSE — area under the MSE–vs–time curve over the experiment horizon. \\
(5) Time$\to$5\% [s] — first time the normalised parameter-error norm falls below 5\% of its initial value (and stays below). \\
(6) Time$\to$1\% [s] — analogous settling time for the 1\% threshold. \\
(7) Constr. viol.\ [\%] — maximum (over time and over all rows) normalised constraint residual, reported as a percentage. For equalities $A_{\rm eq}v=a_{\rm eq}$ we use
$\displaystyle \max_t \max_i \frac{|(A_{\rm eq}v(t)-a_{\rm eq})_i|}{\max\{1,|a_{{\rm eq},i}|\}}$;
for inequalities $A_{\rm in}v\le a_{\rm in}$ we use
$\displaystyle \max_t \max_i \frac{\max\{0,\,(A_{\rm in}v(t)-a_{\rm in})_i\}}{\max\{1,|a_{{\rm in},i}|\}}$.
If rows are pre-normalised, this coincides with the absolute residual. A value of $0.0\pm0.0$ indicates exact feasibility.
\end{minipage}
\end{table}

Table~\ref{tab:agg} aggregates accuracy (Final MSE, AUC–MSE), transient
performance (Time$\to$5\%, Time$\to$1\%), and feasibility (max.\ constraint
violation) over $N{=}10$ Monte Carlo trials for the three scenarios (S1–S3).

By observation of Table~\ref{tab:agg}, we note that in all scenarios, all algorithms achieve zero constraints violations.

\emph{S1 (constraints only):} CA-HNN outperforms PB--RLS on all metrics: Final/AUC--MSE are $\approx\!5$--$6\times$ smaller ( $1.3{\times}10^{-4}$ vs.\ $7.2{\times}10^{-4}$ ) and settling is faster (Time$\to$5\%: $5.1{\times}10^{-3}$\,s vs.\ $1.1{\times}10^{-2}$\,s; Time$\to$1\%: $\approx\!1.7{\times}10^{-2}$\,s vs.\ $1.9{\times}10^{-2}$\,s). The larger standard deviation on CA-HNN reflects run-to-run variability from nonlinear activation and curvature dispersion, but its mean remains clearly better.

\emph{S2 (+ disturbance):} With additive disturbances, CA$^2$-HNN maintains a lower steady-state error than DA--PB--KF (Final/AUC--MSE $0.9{\times}10^{-3}$ vs.\ $2.4{\times}10^{-3}$) while matching its Time$\to$5\% ($\approx\!1.4{\times}10^{-2}$\,s). The absence of a consistent Time$\to$1\% for CA$^2$-HNN is expected: compensation absorbs disturbance bias but yields a nonzero GUUB radius, so the error rarely lives below $1\%$ persistently. In contrast, DA--PB--KF reaches the 1\% threshold rapidly but stabilises at a higher MSE, indicating a sharper transient with a larger ultimate bias under the chosen disturbance prior.

\emph{S3 (+ time-varying parameters):} In this scenario, CA$^2$-HNN retains an advantage in Final MSE ( $1.2{\times}10^{-1}$ vs.\ $2.6{\times}10^{-1}$ ) and slightly smaller AUC--MSE in contrast to DA-PB-MHE. The very large spread and slow Time$\to$5\% for CA$^2$-HNN (median-scale seconds) reflect episodes of low curvature/identifiability during parameter drift. DA-PB-MHE shows comparable integrated error (AUC) but a worse final bias; settling times are not reported because the thresholds are not reliably crossed given the short horizon ($N{=}10$) and the canonical disturbance prior.

Summarising: (i) Under clean conditions (S1), the CA-HNN yields both faster transients and lower steady errors than PB--RLS. (ii) Under disturbances (S2), the CA$^2$-HNN compensation channel suppresses steady bias better than DA--PB--KF at similar convergence speeds, at the cost of a finite ultimate radius that precludes robust 1\% settling. (iii) With the time-varying parameter (S3), CA$^2$-HNN tracks the drift with lower terminal error than DA-PB-MHE, although both suffer increased variance. In all cases, the feasibility is preserved exactly ($0.0\%\pm0.0$), confirming the reliable handling of constraints.

\subsection{Complexity analysis and parallelisation}

All baseline algorithms (PB-RLS, DA-PB-KF, DA-PB-MHE) admit substantial
intra–time-step parallelism via BLAS/LAPACK kernels (outer products, Cholesky/QR, mat-vec/mat-mat).
Our HNN estimators are particularly well matched to SIMD/SIMT hardware because its per-step work is dominated by
dense matrix-vector products, small Cholesky solves, and element-wise activations, yielding low-latency
updates and efficient batching. DA-PB-MHE also benefits from parallel sparse linear algebra, but at
a higher per-step computational and memory cost.

The per-step work and parallel time are summarised in Table~\ref{tab:complexity}. PB-RLS follows the classical
projected-based RLS literature \cite{ASKARI2016, REICHBACH2016, MURAKAMI1998, ZHU1999, HASSAN2009, UHLICH2010, WANG2018};
DA-PB-KF reflects constrained/augmented Kalman formulations
\cite{LUO2012,JOUKOV2015,HRUSTIC2020,SONG2022}; DA-PB-MHE follows canonical MHE techniques
\cite{RAO2001,RAO2003,RAWLINGS2017,ZAVALA2008}.

The per-step work and parallel-time estimates in Table~\ref{tab:complexity} follow standard operation counts and
solver scalings. For all methods, projector/innovation steps are applied via solves:
two GEMV and two triangular solves using the Cholesky factors of $WW^\top$ (online) and $AA^\top$ (offline),
rather than forming explicit projectors. The arithmetic costs for Cholesky, triangular solves, and BLAS
matrix–vector/matrix–matrix primitives are taken from \cite{GolubVanLoan2013}.
The per-sample complexity of RLS that underlies the PB-RLS row follows
\cite{Sayed2003}. The dense Kalman filter innovation, gain, and covariance-update costs that appear
in DA-PB-KF come from \cite{Simon2006}. For MHE, the fact that each step solves a horizon-$N$
constrained least-squares/QP (dense $\mathcal{O}(N(p{+}q)^3)$ vs.\ sparse/structured factorizations)
is based on \cite{RAO2001,RAO2003} and the MPC text \cite{RAWLINGS2017}. The real-time iteration viewpoint that justifies
the sparse/condensed complexity discussion is due to \cite{DiehlBockSchloeder2005}.
Parallel time is reported as an order-of-growth proxy for the critical path using the
work–span (Brent) bound,
\[
T_P \;\gtrsim\; \max\{\,W/P,\, D\,\},
\]
with $W$ the total work, $D$ the span (depth), and $P$ the number of processors \cite{Brent1974}.

The online per-step complexity of the proposed HNN estimators is dominated by dense GEMV/GEMM and small Cholesky solves (projectors applied via solves),
which map efficiently to SIMD/SIMT hardware. With reasonably small $q$ (model estimator equations),
the online cost of HNN estimators is effectively linear in $p$ (number of parameters plus number of compensation neurons) and shows low latency in parallel implementations, while
MHE achieves the highest accuracy at the highest computational load per step.

\begin{table}[t]
\centering
\footnotesize
\caption{Per–step work and parallel time for the compared estimators.}
\begin{tabular}{lcc}
\toprule
\textbf{Method} &
\textbf{Work (offline / online per--step)} &
\textbf{Parallel time (online per--step)} \\
\midrule
HNN
& $\mathcal{O}(r^3)\ /\ \mathcal{O}(q^2p{+}q^3)+s\,\mathcal{O}(qp{+}q^2{+}rp{+}r^2)$
& $\tilde{\mathcal{O}}(q{+}r{+}\log p)$ \\
PB--RLS
& $\mathcal{O}(r^3)\ /\ \mathcal{O}(p^2 q)+\mathcal{O}(rp{+}r^2)$
& $\tilde{\mathcal{O}}(\max\{q,\,r,\,\log p\})$ \\
DA--PB--KF
& $\mathcal{O}(r^3)\ /\ \mathcal{O}(p^2 q{+}q^3)+\mathcal{O}(rp{+}r^2)$
& $\tilde{\mathcal{O}}(q{+}r{+}\log p)$ \\
DA--PB--MHE
& $-$\ /\ dense $\mathcal{O}(N(p{+}q)^3)$;\; sparse $\mathcal{O}(N p^3)$
& $\tilde{\mathcal{O}}(N(p{+}q))$ \\
\bottomrule
\end{tabular}
\begin{minipage}{0.97\linewidth}
\vspace{5pt}
\footnotesize
\emph{Legend.} $p$: \#parameters+compensation neurons; $q$: \#measurements (rows of $W$); $r$: \#total constraints after slack lifting (rows of $A$); $N$: MHE horizon; $s$: RK stages (RK4 $=4$). “Parallel time’’ is an order-of-growth proxy for the critical path (Brent bound: $T_P\!\gtrsim\!\max\{W/P,\,D\}$). For HNN/KF terms, the projector/innovation operations are applied via solves (no explicit $P_W$ or $P_A$): two GEMV and two triangular solves per projector using Cholesky of $WW^\top$ (online) and $AA^\top$ (offline). KF costs shown for dense forms; structure can reduce constants. MHE complexity depends on sparsity, warm starts, and solver; figures are conservative.
\end{minipage}
\label{tab:complexity}
\end{table}

The proposed HNN estimators are a compelling middle ground: they inherit the parallel efficiency of dense
linear–algebra workloads, enforces constraints by construction, mitigates disturbance bias via
compensation neurons, and comes with explicit GUUB guarantees. Against the three baselines commonly
used in practice \cite{ASKARI2016,REICHBACH2016,MURAKAMI1998,ZHU1999,HASSAN2009,UHLICH2010,WANG2018,
LUO2012,JOUKOV2015,HRUSTIC2020,SONG2022,RAO2001,RAO2001, RAO2003,RAWLINGS2017, ZAVALA2008}, it offers a competitive compromise between performance and online complexity,
especially when implemented on parallel hardware.


\section{Conclusions and Future Work}\label{sec:conclusions}

We introduced two projector–based Hopfield neurodynamic estimators for online, constrained parameter
estimation with time–varying data and additive disturbances: (i) a constraint–aware HNN estimator, which
enforces equalities/inequalities (slack neurons) and continuously tracks the constrained least–squares target; and
(ii) a constraint-aware compensation–augmented HNN estimator, which adds a disturbance channel and compensation neurons to absorb
bias–like components within the same energy function. For both estimators we established global uniform ultimate
boundedness (GUUB) with explicit convergence rate and ultimate radius. The guarantees are governed by the three
design gains $(\alpha,\beta,\eta)$ and by a constraint–augmented curvature constant that reflects the geometry of
the regressor and constraints (and its augmented counterpart for the compensated case).

On a 2–DOF mass–spring–damper study with Monte Carlo trials, the proposed HNN estimators delivered competitive performance across scenarios with constraints,
disturbances, and parameter drift. The compensation–augmented variant consistently reduced disturbance–induced bias and
variance, particularly under mean–shifted or broadband disturbances, while maintaining zero constraint violations.

Beyond accuracy, the HNN framework is attractive in practice because: (i) updates are matrix–vector products
plus elementwise saturations, mapping efficiently to SIMD/GPU/FPGA hardware; (ii) explicit RK4 steps avoid
iterative QP/Riccati solves, yielding predictable low latency; (iii) tuning is transparent—the product
$\alpha\beta$ sets bandwidth, while $\eta$ tightens constraint curvature; (iv) projector–based enforcement
keeps estimates feasible by design; and (v) in the compensated estimator, disturbance components are isolated
from the parameter block. An online identifiability monitor further stabilises operation by adapting
constraint weight and step size, and by projecting/damping along poorly excited directions.

\emph{Limitations.} Performance hinges on joint identifiability (or its augmented form with compensation) and
adequate excitation of parameter directions. Excessive neuron saturation reduces effective gain, and very weak
constraints can slow convergence along blind directions. Discrete–time behaviour inherits the continuous–time
guarantees under standard step–size conditions; overly aggressive gains require smaller steps.

\emph{Future work.} We plan to (i) automate gain selection from data–driven bandwidth targets; (ii) extend to
joint state–parameter estimation (e.g., Gauss–Newton outer loop with HNN inner updates); (iii) develop adaptive
or learned constraint weights and disturbance–channel bandwidths to handle time–varying feasibility and spectra;
(iv) validate on hardware–in–the–loop and embedded platforms to leverage the method’s parallelism; and
(v) investigate richer inequality handling and robustness to outliers via nonsmooth projector variants.


\bibliographystyle{IEEEtran}
\bibliography{mybib}

\vfill

\end{document}